\documentclass[aps,pre,preprint,groupedaddress,showpacs]{revtex4-1}
\usepackage[dvips]{graphicx}
\usepackage{textcomp}
\usepackage{amssymb}

\newcommand{\mc}{\multicolumn}

\begin{document}

\title{
\Large\bf Three-dimensional $O(N)$-invariant $\phi^4$ models at criticality for $N\ge 4$}

\author{Martin Hasenbusch}
\affiliation{
Institut f\"ur Theoretische Physik, Universit\"at Heidelberg,
Philosophenweg 19, 69120 Heidelberg, Germany}

\date{\today}
\begin{abstract}
We study the  $O(N)$-invariant $\phi^4$ model on the simple cubic
lattice by using Monte Carlo simulations. By using a finite size scaling 
analysis, we obtain accurate estimates for the critical exponents $\nu$ and 
$\eta$ for $N=4$, $5$, $6$, $8$, $10$, 
and $12$. We study the model for each $N$ for at least three different
values of the parameter $\lambda$ to control leading corrections to scaling. 
We compare 
our results with those obtained by other theoretical methods.
\end{abstract}

\maketitle
\section{Introduction}
We study the $\phi^4$ model on the simple cubic lattice by using 
Monte Carlo (MC) simulations combined with a finite size scaling
(FSS) \cite{Barber} analysis. The $\phi^4$ model
is a prototypical model to investigate critical phenomena. 
The reduced Hamiltonian of the $N$-component $\phi^4$ model, for a vanishing
external field, is given by
\begin{equation}
\label{hamiltonian}
{\cal H} = - \beta \sum_{<x,y>} \vec{\phi}_x \cdot \vec{\phi}_y
   + \sum_{x} \left[\vec{\phi}_x^{\;2} + \lambda (\vec{\phi}_x^{\;2} -1)^2   \right]
\; ,
\end{equation}
where $\vec{\phi}_x$ is a vector with $N$ real components.
We  label the sites of the simple cubic lattice by
$x=(x_0,x_1,x_2)$, where $x_i \in \{1,2,\ldots,L_i\}$. Furthermore,
$<xy>$ denotes a pair of nearest neighbors on the lattice.
For $\lambda=0$ we get the Gaussian model or free field theory on the lattice.
For $\lambda > 0$, the model undergoes a second order phase transition.
In the limit $\lambda \rightarrow \infty$ the field is forced to unit
length $|\vec{\phi}_x| =1$.  In the literature, the model in this limit
is referred to as $O(N)$ vector model. In the following we shall use the 
notation $\lambda=\infty$ to refer to this limit.

The modern theory of critical phenomena is the renormalization group (RG),
going back to the seventies of the last century.
The RG-theory furnishes a general framework but also provides
computational tools like the $\epsilon$-expansion or the functional
renormalization group (FRG) method.  

In the neighborhood of the critical temperature, at a second order phase
transition, thermodynamic quantities diverge, following power laws. For example,
the correlation length $\xi$ in the thermodynamic limit, behaves as
\begin{equation}
\label{powerlawxi}
\xi = a t^{-\nu} \; \left(1 + b  t^{\theta} + c t + ... \right) \;,
\end{equation}
where $t=(T-T_c)/T_c$ is the reduced temperature and $\nu$ the critical 
exponent of the correlation length.  The power law is subject to 
confluent and analytic corrections. In eq.~(\ref{powerlawxi}) we give 
the leading ones. The correction exponent can be written as $\theta=\nu \omega$,
where $\omega$ naturally appears in FSS.

The RG predicts that second order phase transitions fall into universality
classes. For a given class, critical exponents such as $\nu$ and correction
exponents such as $\omega$ assume unique values. The same holds for so called
amplitude ratios.
A universality class is characterized by a few qualitative features
of the system. These are the spatial dimension, the symmetry properties of the
order parameter and the range of the interaction.
Accurate experiments and theoretical calculations support
the universality hypothesis. In the case of the model discussed here, the 
universality class, for a given value of $N$, should not depend on the 
parameter $0 < \lambda \le \infty$.
For reviews on critical phenomena and the RG see, for example 
\cite{WiKo,Fisher74,Fisher98,Kleinert,PeVi02}.

Great progress has been achieved recently by using the so called
conformal bootstrap (CB) method.
In particular in the case of the three-dimensional
Ising universality class, corresponding to $N=1$, 
the accuracy that has been reached for critical
exponents clearly surpasses that of other theoretical methods. See refs.
\cite{Kos:2016ysd,Simmons-Duffin:2016wlq} and references therein. Very recently also highly accurate estimates \cite{che19} were obtained for the XY
universality class, $N=2$,
surpassing the accuracy of results obtained by lattice
methods. In the case of the Heisenberg universality class, $N=3$,
accurate results were provided in \cite{che20}.
Results for $N=4$  are given in refs. \cite{Kos15,CaHaSe16}.
For a review on the CB method, see for example \cite{revCB}.

In the last few years, progress has also been achieved by using other methods.
The $\epsilon$-expansion for critical exponents of
$O(N)$-invariant models has been extended to 6-loops \cite{KoPa17} and
to  7-loops \cite{Sch18}. An analysis of the  7-loop series is provided, for
example, in refs. \cite{Sha21,AbSa21}.
In ref. \cite{DePo20} accurate results were obtained for critical
exponents and the correction exponent $\omega$ for three-dimensional
$O(N)$-invariant systems by using the FRG method. For a recent review 
on the FRG method see ref. \cite{Eich21}.
These two approaches can be applied to a wide range of problems.
For example, both the $\epsilon$-expansion as well as the FRG
method can be applied to dynamic problems.

The Monte Carlo simulation of lattice models in combination with a FSS 
analysis is well established in the study of critical phenomena and 
might serve as benchmark of these methods.
Recently in 
refs. \cite{Landau18,myIso,Xu19,myClock,myIco} accurate estimates of the 
critical exponents for $N=1$, $2$, and $3$ were obtained. 
For references to the  bulk
of previous work see refs. \cite{Landau18,myIso,Xu19,myClock,myIco}
and the review \cite{PeVi02}.
Here we like to bridge the gap to the $1/N$ expansion. To this end we 
study the $\phi^4$ model for $N=4$, $5$, $6$, $8$, $10$, and $12$. 
Previous simulations for these values of $N$ are discussed in the conclusions.
For a review on the large $N$-expansion see, for example \cite{MoZi03}.
For a physical motivation to study the $\phi^4$ model for $N=4$ and $5$, 
see for example sections 6.1 and 6.2 of the review \cite{PeVi02}.

The outline of the paper is the following. In section \ref{phi4model}, 
we define the observables that we measure. In section 
\ref{sec_corrections} we discuss corrections to scaling. In the main
part of the paper, we discuss our simulations and the analysis of the data.
In section \ref{algo} we briefly sketch the Monte Carlo 
algorithm  and its implementation.
In sections \ref{ana4} and \ref{ana5} we discuss the cases $N=4$ and $5$, 
respectively. The simulations and the analysis of the data for $N \ge 6$
are briefly sketched in section \ref{ana6plus}. 
Finally, in section \ref{comparison} we conclude and
compare our results for the critical exponents with those obtained 
by various methods given in the literature.

\section{The observables}
\label{phi4model}
Here we study the same observables as in previous work, see for example section
II B of ref. \cite{myClock}. For completeness we recall the definitions of the
quantities that we have measured.

In our study, the linear lattice size $L=L_0=L_1=L_2$
is equal in all three directions throughout. We employ periodic boundary conditions.

The energy of a given spin configuration is defined as
\begin{equation}
\label{energy}
 E=  \sum_{<xy>}  \vec{\phi}_x  \cdot \vec{\phi}_y \;\;.
\end{equation}

The magnetic susceptibility $\chi$ and the second moment correlation length
$\xi_{2nd}$ are defined as
\begin{equation}
\chi  \equiv  \frac{1}{V}
\left\langle \Big(\sum_x \vec{\phi}_x \Big)^2 \right\rangle \;\;,
\end{equation}
where $V=L^3$ and
\begin{equation}
\xi_{2nd}  \equiv  \sqrt{\frac{\chi/F-1}{4 \sin^2 \pi/L}} \;\;,
\end{equation}
where
\begin{equation}
F  \equiv  \frac{1}{V} \left \langle
\Big|\sum_x \exp\left(i \frac{2 \pi x_k}{L} \right)
        \vec{\phi}_x \Big|^2
\right \rangle
\end{equation}
is the Fourier transform of the correlation function at the lowest
non-zero momentum. In our simulations, we have measured $F$ for the three
directions $k=0,1,2$ and have averaged these three results.

In addition to elementary quantities like the energy, the
magnetization, the specific heat or the magnetic susceptibility, we
compute a number of so-called phenomenological couplings, that means
quantities that, in the critical limit, are invariant under RG
transformations.
We consider the Binder parameter $U_4$ and its sixth-order generalization
$U_6$, defined as
\begin{equation}
U_{2j} \equiv \frac{\langle \vec{m}^{2j}\rangle}{\langle \vec{m}^2\rangle^j},
\end{equation}
where $\vec{m} = \frac{1}{V} \, \sum_x \vec{\phi}_x$ is the
magnetization of a given spin configuration. We also consider the ratio
$R_Z\equiv Z_a/Z_p$ of
the partition function $Z_a$ of a system with anti-periodic boundary
conditions in one of the three directions and the partition function
$Z_p$ of a system with periodic boundary conditions in all directions.
This quantity is computed by using the cluster algorithm.
For a discussion see Appendix A 2 of ref. \cite{XY1}.
In the following we shall refer to the RG-invariant
quantities $U_{2j}$, $Z_a/Z_p$ and $\xi_{2nd}/L$ 
using the symbol $R$.

In our analysis we need the observables as a function of $\beta$ in
some neighborhood of the simulation point. To this end we have
computed the coefficients of the Taylor expansion of the observables
up to the third order.
For example the first derivative of the expectation value
$\langle A \rangle$ of an observable $A$ is given by
\begin{equation}
\frac{\partial \langle A \rangle}{\partial \beta} = \langle A E \rangle
- \langle A \rangle \langle E \rangle \;\;.
\end{equation}

\section{Corrections to scaling}
\label{sec_corrections}
In the analysis of the data corrections to scaling play an important 
role. Systematic errors are caused by corrections that are not or not 
exactly taken into account in the Ans\"atze that are used to fit the data.

Based on the general framework of the RG-theory we expect that the quantities
that we are dealing with in the FSS analysis behave at the critical point as
\begin{equation}
 A(L) = a L^{x} (1 + \sum_i b_i L^{-\omega_i} + 
                 \sum_{ij} c_{ij} L^{-\omega_i-\omega_j} +...) \;,
\end{equation}
where $x$ is the critical exponent we like to determine and $\omega_i>0$
are correction exponents. RG-theory predicts that $x$ and $\omega_i$ are
universal, while $a$, $b_i$ and $c_{ij}$ depend on the particular system 
that is considered. In the case of the model studied here, these 
coefficients are functions of the parameter $\lambda$. 

In practice only a small number of correction terms can be taken into 
account, since the statistical error of the estimates obtained by the 
fit rapidly increases with the number of free parameters. The systematic
error can be reduced by going to larger lattice sizes $L$.
However we are limited in this direction, since the CPU time that is required
to keep the statistical error constant  essentially grows as $L^{d+z}$, 
where $d$ is the dimension of the system and the dynamical critical exponent
$z>0$ for the algorithms used here.
In order to decide on the design of the study, information 
on the corrections is needed. Getting information by using FSS studies, 
in practice one is restricted to the leading correction. In the 
case of subleading corrections we have to rely on other theoretical methods. 

Various methods give consistently $\omega \approx 0.8$ for the exponent of the
leading correction to scaling for $N \lessapprox 12$.
Since the amplitudes of corrections $b_i$ depend on the details of the system,
in our case, one might try to find a value $\lambda^*$ such
that the amplitude of the leading correction vanishes:
$b(\lambda^*)=0$. RG-theory tells us that $\lambda^*$ is the same for 
different quantities.  Models with a vanishing amplitude of the leading 
correction to scaling are denoted as improved models. 
This idea had been exploited first by using high temperature series
expansions of such models \cite{ChFiNi,FiCh}. For early  Monte Carlo
simulations of improved models sharing the universality class of the
three-dimensional Ising model see for example refs.
\cite{Bloete,Ballesteros,KlausStefano}.
In \cite{CaPe99}
it has been pointed out that for the $\phi^4$ model on the simple cubic 
lattice in the large $N$ limit $\lambda^*$ does not exist. For $N=1$, 
$2$, $3$, and $4$ a finite  $\lambda^*$ has been found. Most recent 
estimates are $\lambda^* \approx 1.1$ \cite{myPhi4} for $N=1$, 
$\lambda^*=2.15(5)$ \cite{XY2} for $N=2$, $\lambda^*=5.17(11)$  
\cite{myIco} for $N=3$ 
and $\lambda^*=20_{-6}^{+15}$ \cite{O234} for $N=4$. The value of $\lambda^*$
is rapidly increasing with $N$. In the case of $N=5$ it is not fully 
settled, whether $\lambda^*$ exist. If yes, it is close to the limiting 
case $\lambda=\infty$ \cite{ourO5}. 
Analyzing our data, we confirm that $\lambda^*$ for $N=4$ exists, 
while for $N=5$ this is highly unlikely. The result that for $N>5$ 
no $\lambda^*$ exists is very robust.  
As a consequence, the outline of the study for $N=4$ is very similar 
to our recent work \cite{myClock,myIco}, where we studied models in the 
XY and Heisenberg universality classes. For larger values of $N$ we focus
on improved observables, which are constructed such that leading corrections
are suppressed for any model. Still the accuracy of the estimates of critical
exponents is lower for larger values of $N$ than for $N=4$. 

\subsection{Subleading corrections}
\label{sec_subleading}
In the analysis of our data, we use  prior information on subleading 
corrections to scaling. In section III A of ref. \cite{myClock} we argue,
based on the literature, that there should be only a small
dependence of the irrelevant RG-eigenvalues on $N$. Therefore
the discussion of section III A of ref. \cite{myClock} should
apply to the present case $4 \le N \le 12$ at least on a qualitative level.

The most important subleading correction should be
due to the breaking of the rotational symmetry by the simple
cubic lattice. Corrections related with the spatial anisotropy
are discussed in ref. \cite{ROT98}. To this end, the two-point
function of $O(N)$-invariant models is studied by using the 
$1/N$-expansion, field-theoretic methods and the high temperature  series
expansion. Results for $\sigma$, for various values of $N$,
are summarized in table VI 
(table VIII of the preprint version), where the correction exponent is
$\omega_{NR} = 2 + \sigma$.
In the large $N$-limit one obtains \cite{ROT98}
\begin{equation}
 \sigma = \frac{32}{21 \pi^2 N} + O\left(N^{-2} \right)  \;.
\end{equation}
According to the authors, this expression gives a reasonable numerical
value at best down to $N=8$. Looking at table VI of ref. \cite{ROT98},
it seems plausible
that $\sigma$ has a plateau-like maximum at $N=3$ up to $4$ and then slowly
decreases.  This behavior is supported both by the high temperature 
series expansion as well as the field-theoretic methods. 

We might gauge the estimates of ref. \cite{ROT98} by using the highly 
accurate result $\sigma=0.022665(28)$ obtained by using the CB method 
in ref. \cite{Simmons-Duffin:2016wlq} for $N=1$.  
This suggests in particular that
for the values of $N$ studied here, the field theoretic estimates of 
$\sigma$ given in ref. \cite{ROT98} are too small.

Based on these considerations, we use the numerical value
\begin{equation}
\label{oursigma}
 \sigma=\mbox{min}\left[0.023,\frac{32}{21 \pi^2 N}\right]
\end{equation}
with an error of $0.005$ in the analysis of our data. We checked that
the estimates of the quantities we are interested in change only by little, 
when $\sigma$ is varied within this error band.

In addition, there are corrections that are intrinsic to the quantity that
is studied. For example in the case of the magnetic susceptibility there 
is the analytic background. This can be interpreted as a correction with the
exponent $2 - \eta$.  It should also appear in the Binder cumulant $U_4$ that
contains $\langle m^2 \rangle$ in its definition. In the case of $\xi_{2nd}/L$
there is, by construction, a correction with the exponent $2$. 

\section{The Monte Carlo algorithm and its implementation}
\label{algo}
We simulated by using a hybrid of local updates and the 
wall cluster algorithm \cite{KlausStefano}. The probability to delete the 
link between the nearest neighbor sites $x$ and $y$ is the same as for 
the Swendsen-Wang (SW)
\cite{SW} or the single  cluster algorithm \cite{Wolff}. It differs in the 
choice of the clusters to be flipped. In the case of the SW algorithm, all
clusters are constructed, and a cluster is flipped with probability $1/2$.
In the case of the single cluster algorithm, a single site of the lattice 
is randomly chosen.  The cluster that contains this site is flipped. 
To this end, only this cluster has to be constructed. In the case of the 
wall cluster algorithm, a plain of the lattice is randomly selected.
All clusters that share a site with this plain are flipped. Also here, only 
these clusters need to be constructed.  In order to determine $Z_a/Z_p$, 
we have to go through all $N$ components of the field \cite{XY1}.

We have implemented overrelaxation updates 
\begin{equation}
 \vec{\phi}_x^{\;\;'} =
 2 \frac{\vec{\Phi}_x \cdot \vec{\phi}_x}{\vec{\Phi}_x^2} \vec{\Phi}_x 
- \vec{\phi}_x \;\;,
 \end{equation}
 where
\begin{equation}
\vec{\Phi}_x = \sum_{y.nn.x}  \vec{\phi}_y \;\;,
\end{equation}
where $\sum_{y.nn.x}$ is the sum over all nearest neighbors $y$ of the 
site $x$. 
Note that these updates do not change the value of the Hamiltonian and
therefore no
accept/reject step is needed. It is computationally quite cheap, since
no random number and no evaluation of $\exp()$ is needed.
In the case of the
overrelaxation update we run through the lattice in typewriter fashion.
As the cluster update, the overrelaxation update does not change 
$|\vec{\phi}_x|$.

For finite  $\lambda$, we perform Metropolis updates to change 
$|\vec{\phi}_x|$. For each site, 
we perform two subsequent updates. We use the acceptance probability
\begin{equation}
 P_{acc} = \mbox{min}\left[1,\exp(-H(\phi') + H(\phi))\right]  \;.
\end{equation}
For the first hit, we generate the proposal by
\begin{equation}
 \phi_{x,i}' =  \phi_{x,i} + s_1 (r -0.5) 
\end{equation}
for each component $i$ of the field at the site $x$. $r$ is a uniformly distributed 
random number in $[0,1)$ and the step size $s_1$ is tuned such that the acceptance rate
is roughly $50 \%$. In the case of the second hit, we randomly select a single component
$j$. Now $\phi_{x,j}' =  \phi_{x,j} + s_2 (r -0.5)$, while all other components keep their
value. Also here, we tune $s_2$ such that the acceptance rate is roughly $50 \%$. Also here, we run through the lattice in typewriter fashion.

In the limit $\lambda=\infty$, we simulated the model by using a hybrid of 
the overrelaxation algorithm and the wall cluster algorithm \cite{KlausStefano}.
Below we give the update sequence used for $\lambda=\infty$ and $N=5$, $6$, $8$, $10$, and $12$ as 
C-code: \\
\verb| ROTATE; over(); over(); for(ic=0;ic<N;ic++) wall_0(ic); measure(); | \\ 
\verb| ROTATE; over(); over(); for(ic=0;ic<N;ic++) wall_1(ic); measure(); | \\
\verb| ROTATE; over(); over(); for(ic=0;ic<N;ic++) wall_2(ic); measure(); | \\

Here \verb+over()+ is a full sweep with the overrelaxation update over the 
lattice.
\verb+wall_k(ic)+ is a wall-cluster update with a plain perpendicular to the 
\verb+k+-axis. The component \verb+ic+ of the field is updated. 
The position of the plain
on the \verb+k+-axis is randomly chosen for each component of the field.
Note that in the cluster and Metropolis updates the axis play a special role.
This does not invalidate the updates but might lead to a certain degradation
of the performance. Therefore we interleave the updates with global rotations
of the field. The rotations \verb| ROTATE | 
are build from a sequence of rotations by a random angle between two axis.
For $\lambda=\infty$ the condition $|\vec{\phi}_x^{\;2}|=1$ might be lost due 
to rounding errors. Therefore we normalize the field $\vec{\phi}_x$ after 
each update cycle.
For finite $\lambda$, we have added a sweep with the local two hit Metropolis 
update following \verb|ROTATE|.

In the case of $N=4$ the update cycle for $\lambda=\infty$ is given by \\
\verb| ROTATE; over(); for(ic=0;ic<N;ic++) wall_0(ic); measure(); | \\ 
\verb| over(); for(ic=0;ic<N;ic++) wall_1(ic); measure(); | \\
\verb| over(); for(ic=0;ic<N;ic++) wall_2(ic); measure(); | \\

Again, for finite $\lambda$ a sweep using the Metropolis algorithm is added
for each measurement. 

Note that the composition of the update cycles is not tuned. Essentially
it is based on an ad hoc decision guided by the experience gained in 
previous work.

For a certain fraction of the simulations for $N=4$ we have used the 
SIMD-oriented Fast Mersenne Twister (SMFT) \cite{twister}
pseudo-random number generator, where SIMD is the abbreviation for 
single instruction, multiple data.
In the remaining part of the simulations for $N=4$ and
for larger values of $N$ we have used a hybrid of generators, where 
one component is the \verb$ xoshiro256+ $ taken from \cite{VignaWWW}.
For a discussion of the generator see \cite{ViBl18}. 
As second component we used
a 96 bit linear congruential generator with the multiplier and the
increment $a=c=$\verb+0xc580cadd754f7336d2eaa27d + and the modulus $m=2^{96}$
suggested by O'Neill \cite{ONeill_minimal}. In this case we used our own
implementation. The third component is a multiply-with-carry generator taken 
from \cite{KISS_wiki}.
For a more detailed discussion see the Appendix A of ref. \cite{myIso}. 

Throughout this work, least square fits were performed by using the function
\verb+curve_fit()+
contained in the SciPy  library \cite{pythonSciPy}.
Plots were generated by using the Matplotlib library \cite{plotting}.

Fitting our data, we take lattice sizes $L \ge L_{min}$ into account. 
For small values of $L_{min}$, $\chi^2/$DOF decreases with increasing 
$L_{min}$, since the 
magnitudes of corrections that are not taken into account in the Ansatz
decrease with increasing lattice size $L$. At some point $\chi^2/$DOF levels 
off, since the magnitudes of these corrections become smaller than 
the statistical error. On the other hand, with increasing $L_{min}$, the 
statistical error of the estimates of fit parameters is increasing. Often in 
the literature, the estimates of fit parameters obtained for the 
smallest $L_{min}$ with an acceptable goodness of the fit are taken as the
final results. 

Here, in order to get a better handle on systematic errors due to corrections
that are not included in the Ansatz, 
the final results are chosen such that they are compatible with estimates 
obtained by using several different Ans\"atze, containing more or less 
correction terms.
For a more comprehensive discussion of this issue see section V of 
ref. \cite{myIco}.

\section{The simulations and the analysis of the data for $N=4$}
\label{ana4}
We simulated at 
$\lambda =2$, $4$, $12.5$, $18.5$, $20$, and $\infty$. Let us briefly 
summarize the lattice sizes and the statistics of the simulations. Note
that some of the simulations were already performed a few years ago, leading 
to different choices of the lattice sizes for different values of 
$\lambda$. Most of the simulations were performed on desktop PCs at the
institute of theoretical physics of the university of Heidelberg.
The CPU times quoted below refer to the time that would be needed on 
a single core of an AMD EPYC$^{TM}$ 7351P CPU. For example, on a 
single core of an Intel(R) Xeon(R) CPU E3-1225 v3 the performance of 
our code is very similar.

For $\lambda=\infty$ we simulated the linear lattice sizes
$L=6$, $7$, $8$, ..., $20$, $22$, $24$, ..., $32$, $36$, $40$, $44$, $48$, 
$56$, $64$, $72$, $80$, and $200$. 
Up to $L=20$ we performed $3 \times 10^9$ measurements. For 
$L=22$, $24$, $26$, and $28$ we performed $6.2 \times 10^9$, $5.8 \times 10^9$, 
$3.8 \times 10^9$, and $4.6 \times 10^9$ measurements, respectively. Then
the number of measurements monotonically decreases to $4.8 \times 10^8$ for
$L=80$. For $L=200$ only $7.5  \times 10^6$ measurements were performed.
In total these simulations took about 8.5 years of CPU time.

For $\lambda=20$  we simulated the linear lattice sizes
$L=6$, $7$, $8$, ..., $20$, $22$, $24$, $26$, $30$, $34$, $40$, $50$, $60$, 
$80$, $100$, $140$, $200$, and $300$.  Up to $L=30$ we performed 
$3 \times 10^9$ measurements. For larger linear lattice sizes $L$, the number 
of measurements decreases with increasing $L$. For example for 
$L=200$ and $300$, we performed $9 \times 10^7$ and  $3 \times 10^7$, 
respectively. In total these simulations took about 20.6 years of CPU time.

For $\lambda=18.5$ we simulated the linear lattice sizes
$L=8$, $9$, $10$, ..., $22$, $24$, $26$, ..., $32$, $36$, $40$, $44$, $48$, 
$56$, $64$, ..., $80$, $100$, $120$, and $200$. For $L=10$, we performed 
$5.4 \times 10^9$ measurements. This number slowly drops to $2.8 \times 10^9$
for $L=36$. Then the number of measurements decreases more rapidly with
$L$. 
For example
for $L=80$, $100$, $120$, and $200$, we performed $5.9 \times 10^8$, 
$2.1 \times 10^8$, $2.5 \times 10^8$, and $5.3 \times 10^7$ measurements,
respectively.
In total these simulations took about 19.5 years of CPU time.

For $\lambda=12.5$ we simulated the linear lattice sizes
$L=6$, $7$, $8$, ..., $20$, $24$, $28$, ..., $40$, $48$, $56$, $60$, $64$, and
$80$. Up to $L=28$ we performed $3 \times 10^9$ measurements. 
 For example
for $L=64$ and $80$, we performed $4.6 \times 10^8$ and $1.7  \times 10^8$
measurements, respectively.
In total these simulations took about 7.5 years of CPU time.

For  $\lambda=4$ we simulated the linear lattice sizes
$L=6$, $7$, $8$, ..., $22$, $24$, $26$, $28$, $32$, $36$, ..., $52$, $60$, 
 $70$, $80$. Up to $L=21$ we performed at least $2.4 \times 10^9$ 
measurements. Then the number of measurements decreases with increasing
$L$. For example, for $L=70$ and $80$, we performed $4.8  \times 10^8$
and $1.8  \times 10^8$ measurements, respectively.
In total these simulations took about 6 years of CPU time.

For $\lambda=2$ we simulated the linear lattice sizes
$L=6$, $7$, $8$, ..., $22$, $24$, $26$, $28$, $32$, $36$, ..., $48$, $60$,
$80$. The number of measurements for each lattice size is similar to that
for $\lambda=4$.
In total these simulations took about 4.3 years of CPU time.

Throughout, while the  number of measurements decreases with increasing 
$L$, the CPU time used for a given lattice size $L$, increases with 
increasing $L$. The same holds for larger values of $N$, discussed below.

\subsection{The dimensionless quantities}
First we analyzed the behavior of dimensionless quantities.
We performed joint fits of all four quantities, $Z_a/Z_p$, $\xi_{2nd}/L$,
$U_4$, and $U_6$, for two sets of 
$\lambda$-values. The first set contains  $\lambda=4.0$,
$12.5$, $18.5$, $20$ and $\infty$, while in the second we consider 
$\lambda=2.0$, in addition. We use Ans\"atze of the form
\begin{equation}
\label{masterdimless}
 R_i(\beta_c,\lambda,L) = R_i^*  + 
 \sum_{k=1}^{k_{max}} c_{i,k} [b(\lambda) a_i L^{-\omega}]^k 
  + \sum_j  e_{i,j}(\lambda) L^{-\epsilon_j} \;,
\end{equation}
where $c_{i,1}=1$ and $b(\lambda)$ is normalized such that $a_{Z_a/Z_p}=1$. 
In our fits, we consider $\omega$ as free parameter, while 
we fix the exponents of subleading corrections. In particular
we take $\epsilon_1=2-\eta$, where we took the preliminary estimate 
$\eta=0.03625$, which is close  to our final estimate, eq.~(\ref{etafinalO4}),
$\epsilon_2=2$, and $\epsilon_3=\omega_{NR} = 2.023$, eq.~(\ref{oursigma}). 
The correction with the exponent $\epsilon_1$ 
applies to $U_4$, $U_6$ and $\xi_{2nd}/L$, the correction with $\epsilon_2$
to $\xi_{2nd}/L$, while the correction with the exponent $\omega_{NR}$ is 
non-vanishing in all four cases. The amplitude of the 
leading correction
$b(\lambda)$ is taken as free parameter for each value of $\lambda$.  In 
principle the $e_{i,j}$ depend on $\lambda$. In order to keep the fits 
tractable, we used a parameterization to reduce the number of free parameters.
In the case of $e_{i,1}$, we used
\begin{equation}
\label{parac0}
 e_{i,1}  = d_i \;,
\end{equation}
\begin{equation}
\label{parac1}
 e_{i,1}  = d_i + s_i \lambda^{-1}
\end{equation}
or 
\begin{equation}
\label{parac2}
 e_{i,1}  = d_i + s_i \lambda^{-1}  + t_i \lambda^{-2} \;,
\end{equation}
where $d_i$, $s_i$ and $t_i$ are the free parameters of the fit. In our fits,
$e_{i,2}$ and $e_{i,3}$ are assumed to be constant. For
$e_{i,3}$ this should indeed be a good approximation. In the case of the 
Ising universality class, for the Blume Capel model on the simple cubic 
lattice, in ref. \cite{myIso},
we found that the amplitude of deviations from the rotational invariance 
depends very little on the parameter $D$, where $D$ plays a similar role 
as the parameter $\lambda$ of the model studied here.

In our analysis of the data set 1, i.e. for $\lambda=4.0$,
$12.5$, $18.5$, $20$ and $\infty$, we consider the following choices,
based on the Ansatz~(\ref{masterdimless}): 
\begin{itemize}
\item
Fit 1: $k_{max}=1$, parameterization~(\ref{parac0})
\item
Fit 2: $k_{max}=1$, parameterization~(\ref{parac1})
\item
Fit 3: $k_{max}=2$, parameterization~(\ref{parac0})
\item
Fit 4: $k_{max}=2$, parameterization~(\ref{parac1})
\end{itemize}

\begin{figure}
\begin{center}
\includegraphics[width=14.5cm]{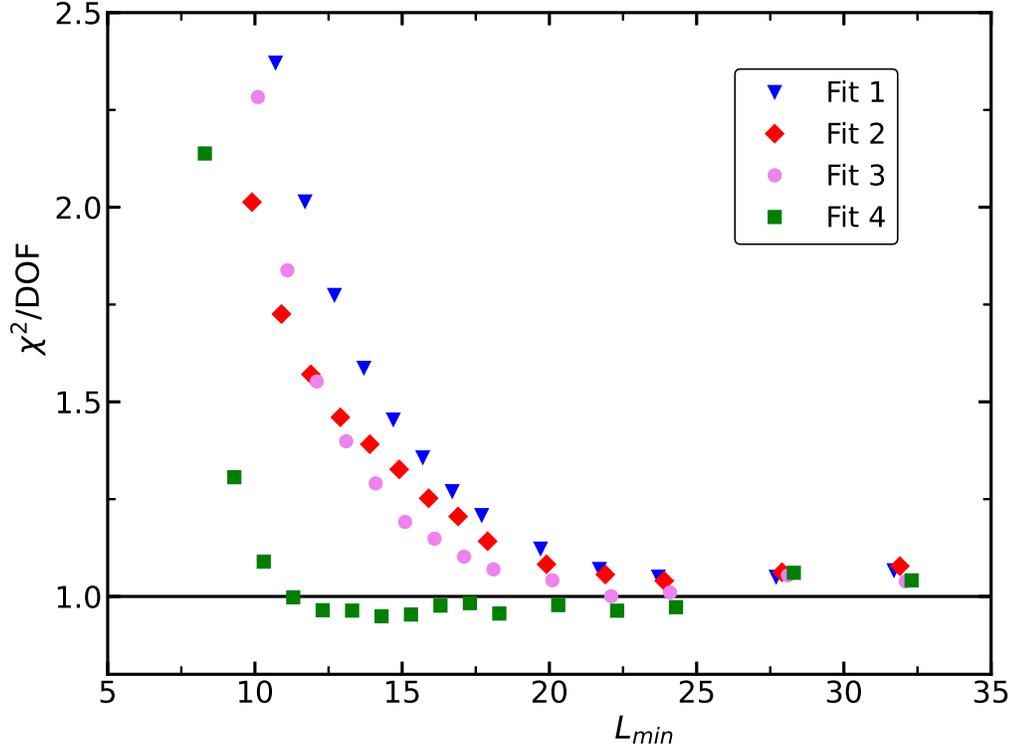}
\caption{\label{ChiDOF_O4}
Joint fits of dimensionless quantities for $\lambda=4.0$, $12.5$, $18.5$, 
$20$, and $\infty$ for $N=4$. 
Numerical estimates for $\chi^2/$DOF obtained from the fits 1, 2, 3, and 4,
which are discussed in the text, are plotted versus the minimal lattice 
size $L_{min}$ that is taken into account.
Note that the values on the $x$-axis are slightly shifted to reduce overlap
of the symbols.
}
\end{center}
\end{figure}

In Fig. \ref{ChiDOF_O4} 
we give the $\chi^2/$DOF obtained in these fits as a function
of the minimal lattice size $L_{min}$ that is taken into account. 
 In the case of 
fit 1, we get $\chi^2/$DOF $=1.07$ corresponding to $p=0.20$ at $L_{min}=22$. 
In the case of fit 2 and fit 3, $\chi^2/$DOF $\approx 1$ is reached for 
somewhat smaller $L_{min}$.  For fit 4, we get $\chi^2/$DOF $=1.00$ already for
$L_{min}=11$. We checked that using the parameterization~(\ref{parac2}) or
adding a term proportional to $L^{-3 \omega}$ improves the goodness of 
the fits only by little.
Finally, we performed fits with an additional correction term on top of fit 4, 
where the correction exponent is a free parameter. Here we get 
$\chi^2/$DOF $=1.06$ and $\chi^2/$DOF $=0.97$ already for $L_{min}=7$ and 
$8$, respectively.
We get $\omega' =4.91(24)$ and $4.88(47)$, for 
$L_{min}=7$ and $8$, respectively. Going to larger $L_{min}$, the 
statistical error of $\omega'$ rapidly increases. The amplitude of this 
correction is comparatively large. Since $\omega'$ is large, 
this correction plays virtually no role for $L \gtrapprox 10$. 
One should note that this finding does not mean that there is no correction
with $2 < \omega' < 4.9$. Such corrections might just have a small amplitude 
compared with the $\omega' \approx 4.9$ correction. 

In a second set of fits we have analyzed in addition the data for $\lambda=2.0$.
Still fit 4 gives $\chi^2/$DOF $=1.07$ corresponding to $p=0.14$ for 
$L_{min}=11$ and $\chi^2/$DOF $=1.02$ corresponding to $p=0.39$ for 
$L_{min}=12$. Using the parameterization~(\ref{parac2}) or
adding a term proportional to $L^{-3 \omega}$ improves the goodness of
the fits only by little.

\begin{figure}
\begin{center}
\includegraphics[width=14.5cm]{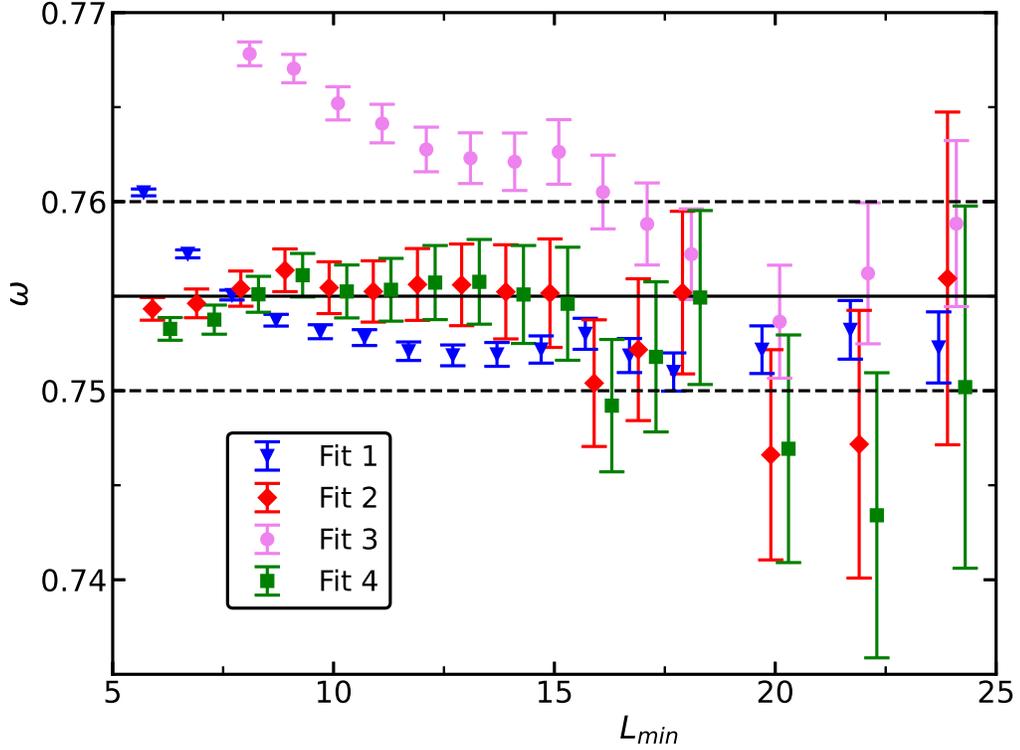}
\caption{\label{Omega_O4}  
Numerical estimates of the correction exponent $\omega$ for $N=4$ obtained 
from joint fits of the data for $\lambda=2$, $4$, $12.5$, $18.5$, $20$, and 
$\infty$ versus the minimal linear lattice size $L_{min}$ that is taken into
account.
The Ans\"atze used in fits 1, 2, 3, and 4 are discussed in 
the text. 
Note that the values on the $x$-axis are slightly shifted to reduce overlap
of the symbols. The solid line gives our final estimate of
$\omega$, while the dashed lines indicate the error.
}
\end{center}
\end{figure}

In Fig. \ref{Omega_O4} we give our results for the correction 
exponent $\omega$ obtained by using the fits 1, 2, 3, and 4 with the data
for $\lambda=2$, $4$, $12.5$, $18.5$, $20$ and $\infty$. 
Here we give all estimates,
irrespective of the  $\chi^2/$DOF. As our final result we consider 
\begin{equation}
 \omega=0.755(5) \;.
\end{equation}
It is chosen such that it contains the results obtained by using the 
fits 2 and 4 up to $L_{min}=15$, while the results of fit 3 are contained for
$L_{min}=18$, $20$, and $22$. The results for fit 1 are contained from 
$L_{min}=7$ up to
$28$. Here and in the following we mean by ``the fit is contained''
that the central estimate obtained by the fit $\pm$ its error lies within 
the interval given by our final result $\pm$ our final error estimate.
To get an idea on the amplitude of corrections to scaling
we quote the results for fit 4 and $L_{min}=12$: 
$b = 0.00474(11)$, $0.00038(11)$, $0.00003(11)$, $-0.00217(11)$, 
$-0.01495(15)$,
and $-0.02868(22)$ for $\lambda=\infty$, $20$, $18.5$, $12.5$, $4.0$, and $2.0$,
respectively.  Taking into account also the results of other fits, 
linearly interpolating  $b(\lambda)$ for $\lambda=18.5$ and $20$ we 
determine the zero of $b(\lambda)$ as
\begin{equation}
\lambda^*=18.4(9) \;,
\end{equation}
which can be compared with the previous estimate $\lambda^*=20_{-6}^{+15}$
\cite{O234}.

\begin{figure}
\begin{center}
\includegraphics[width=14.5cm]{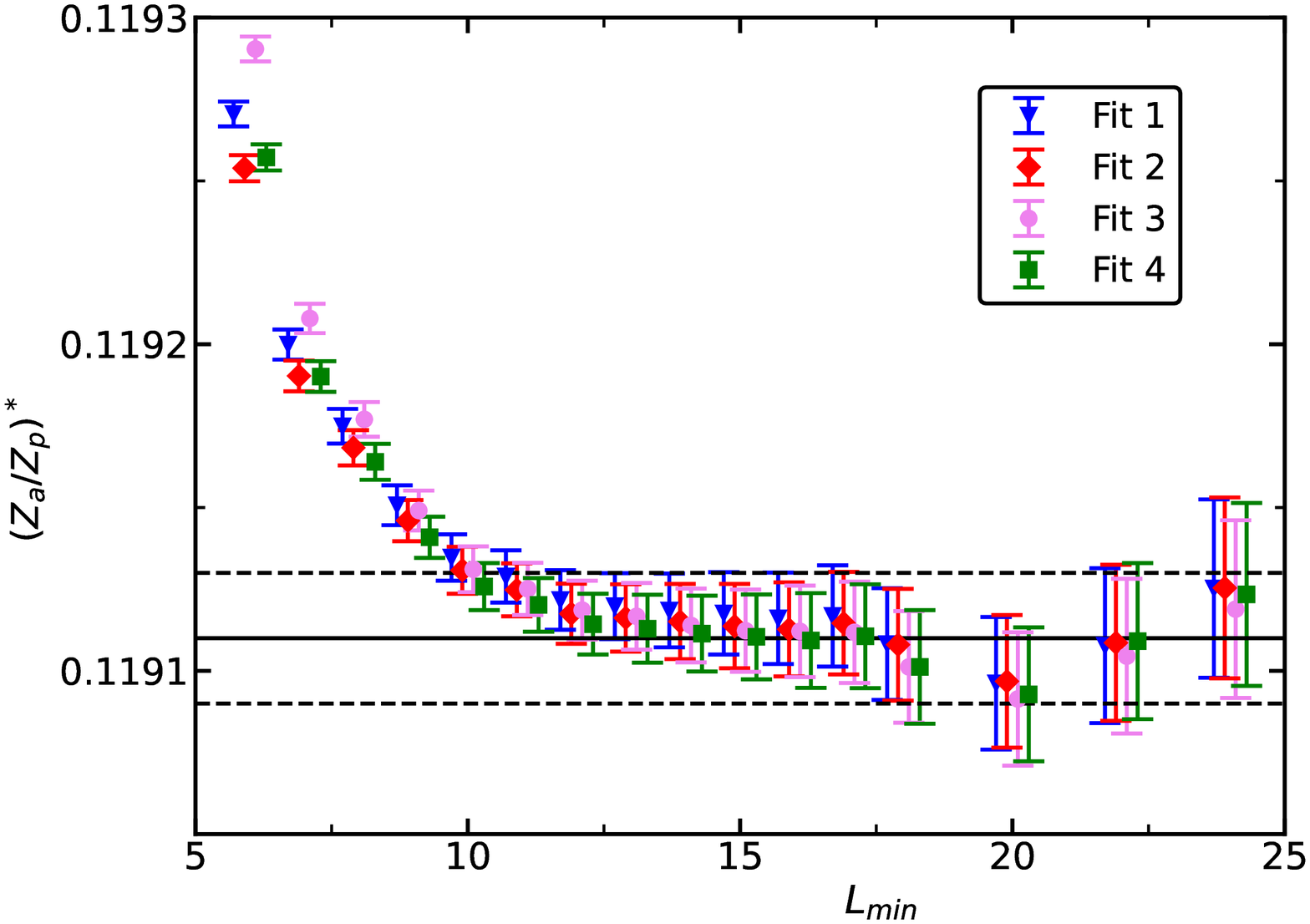}
\caption{\label{ZaZp_O4} 
Data set 1, i.e. $\lambda=4$, $12.5$, $18.5$, $20$, and $\infty$ 
for $N=4$. 
Numerical estimates of $(Z_a/Z_p)^*$ obtained from the fits 1, 2, 3, 4, which
are discussed in the text, are plotted versus the
minimal linear lattice size $L_{min}$ that is taken into account.
The solid line gives our final estimate of 
$(Z_a/Z_p)^*$, while the dashed lines indicate the error.
Note that the values on the $x$-axis are slightly shifted to reduce overlap
of the symbols.
}
\end{center}
\end{figure}

In Fig. \ref{ZaZp_O4} we give our results for the fixed point value  
$(Z_a/Z_p)^*$ of the ratio of partition functions using set 1. 
In comparison with $\omega$, 
the results for $(Z_a/Z_p)^*$ show little dependance on the Ansatz that is
used. The final estimate and its error are chosen such that the results
of all four fits from $L_{min}=12$ up to $17$ are covered.  
Fitting set 2, we get consistent results. 
The final estimates and the errors of the other dimensionless quantities
are determined in a similar way. We get
\begin{eqnarray}
(Z_a/Z_p)^* &=& 0.11911(2) \;, \label{ZaZpstarO4} \\
(\xi_{2nd}/L)^* &=& 0.547296(26) \;, \label{xi2LstarO4} \\
U_4^*   &=& 1.094016(12) \;, \label{U4starO4} \\
U_6^*   &=& 1.281633(33)  \label{U6starO4} \;.
\end{eqnarray}

Finally, in table \ref{betacN4} we summarize the estimates
of the critical temperature obtained in these fits.
Our result for $\lambda=\infty$ is fully consistent
with $\beta_c=0.935856(2)$ given in ref. \cite{Deng06}.

\begin{table}
\caption{\sl \label{betacN4}
Estimates of the inverse critical temperature $\beta_c$ for $N=4$. 
}
\begin{center}
\begin{tabular}{ll}
\hline
 \mc{1}{c}{$\lambda$} &  \mc{1}{c}{$\beta_c$}  \\
\hline
$2.0$     &   0.7978640(4) \\
$4.0$     &   0.85875410(35) \\   
$12.5$    &   0.90951811(21) \\
$18.5$    &   0.91787555(17) \\
$20$      &   0.91919685(15) \\
$\infty$  &   0.93585450(25) \\
\hline
\end{tabular}
\end{center}
\end{table}

\subsection{The magnetic susceptibility and the critical exponent $\eta$}
The magnetic susceptibility at criticality  behaves as 
\begin{equation}
\label{chi1}
\chi=a L^{2-\eta} \;\left(1 + c L^{-\omega} + ...+ d L^{-\omega_{NR}}+ ...\right) 
 + b \;,
\end{equation}
where $b$ is the analytic background. Corrections $\propto L^{-n \omega}$ with 
$n>1$ and further subleading corrections are not explicitly given.
In order to enforce criticality, we take $\chi$ at a fixed value of either 
$Z_a/Z_p$ or $\xi_{2nd}/L$. To this end we take  the fixed point values 
given in eqs.~(\ref{ZaZpstarO4},\ref{xi2LstarO4}). In the following we denote
$\chi$ at a fixed value of $Z_a/Z_p$ or $\xi_{2nd}/L$ by $\bar{\chi}$. 
In the case of fixing $\xi_{2nd}/L$, there is, compared with eq.~(\ref{chi1}),
an additional correction with an exponent equal to $2$.

We consider the improved susceptibility
\begin{equation}
\label{chiimp}
 \bar{\chi}_{imp} = \bar{U}_4^{x} \bar{\chi} \;\;,
\end{equation}
where the exponent $x$ is tuned such that the leading correction to scaling 
is eliminated. In previous work, we determined $x$ in a preliminary analysis, 
and then performed fits by using Ans\"atze based on eq.~(\ref{chi1}) with 
a fixed value of $x$.  Here we perform fits with $x$ as free parameter. In  
appendix \ref{strangefit}, we discuss how we deal with the fact that $x$ 
appears on the left side of the equation.

In a first step we performed joint fits for $\lambda=4.0$, $12.5$, $18.5$, 
$20$, and $\infty$, of $U_4^{x} \chi$ at $\xi_{2nd}/L=0.547296$ or 
$Z_a/Z_p=0.11911$ fixed, where $x$ is a free parameter.
In the case of $\chi$ at $Z_a/Z_p=0.11911$ we used the Ansatz
\begin{equation}
\label{chiansatz1}
 \bar{\chi}_{imp} = a L^{2-\eta} (1+c L^{-\omega_{NR}}) + b \;,
\end{equation}
where $a$ and $b$ are free parameters for each value of $\lambda$ separately, 
while $c$ is the same for all values of $\lambda$.
At $L_{min}=10$ we get $\chi^2/$DOF$=1.02$. The value that we obtain for 
$x$ is stable with increasing $L_{min}$. For example, we get 
$x=-1.67(3)$, $-1.70(3)$, $-1.71(5)$, $-1.72(6)$, and $-1.71(8)$ for 
$L_{min}=10$, $12$, $14$, $16$, and $18$, respectively.  In the following, 
constructing $\bar \chi$ we use $x=-1.7$. 

In the case of $\chi$ at $\xi_{2nd}/L=0.547296$ we used the Ansatz
\begin{equation}
\label{chiansatz2}
 \bar{\chi}_{imp} = a L^{2-\eta} (1+c L^{-2}+d L^{-\omega_{NR}}) + b \;,
\end{equation}
where $a$ and $b$ are free parameters for each value of $\lambda$ separately, 
while $c$ and $d$ are the same for all values of $\lambda$.
Here we find that $\bar{\chi}$, by chance, is only little affected 
by leading corrections.  We find $x \approx -0.07$. 

\begin{figure}
\begin{center}
\includegraphics[width=14.5cm]{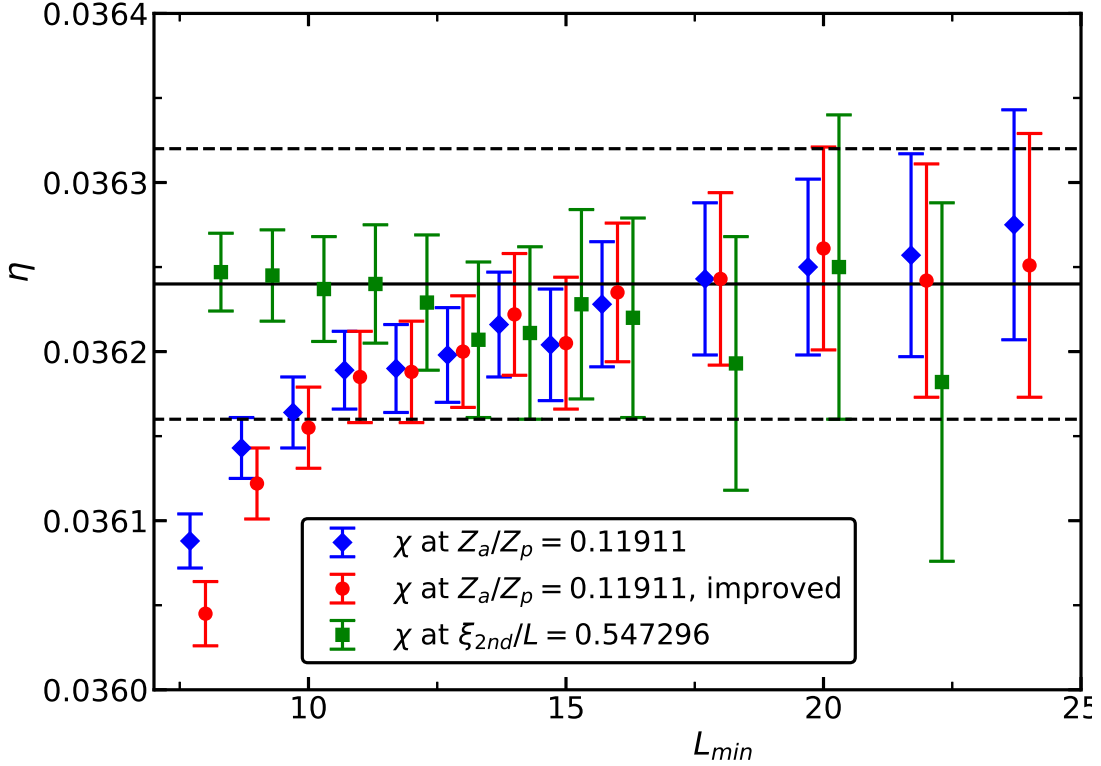}
\caption{\label{Chi_eta_O4}
Estimates of $\eta$ from
joint fits for $\lambda=18.5$ and $20$ for $N=4$. We give the results
of fits for $\chi$ at $Z_a/Z_p=0.11911$ and its improved version obtained 
by using the Ansatz~(\ref{chiansatz1}). Furthermore we give 
the estimates obtained by fitting $\chi$ at $\xi_{2nd}/L=0.547296$ by using 
the Ansatz~(\ref{chiansatz2}). 
Note that the values on the $x$-axis are slightly shifted to reduce overlap
of the symbols.
The solid line gives our final estimate
and the dashed lines indicate the error.
}
\end{center}
\end{figure}

To get the final estimate of $\eta$, 
we have fitted the data for $\lambda=18.5$ and $20$ jointly.  To this 
end, we use the Ans\"atze~(\ref{chiansatz1},\ref{chiansatz2}), but now 
fixing the value of $x$.
In Fig. \ref{Chi_eta_O4} we give the estimates of $\eta$ as a function
of the minimal lattice size $L_{min}$ that is taken into account in the
fits. We have fitted $\chi$ at $Z_a/Z_p=0.11911$ and its improved version
by using the Ansatz~(\ref{chiansatz1}). 
In both cases, we find $\chi^2/$DOF $\approx 1$ and correspondingly
an acceptable  $p$-value for $L_{min} \ge 10$. Note that here and in the 
following we consider $0.1 \lesssim p \lesssim 0.9$ as acceptable.
Instead, $\chi$ at  
$\xi_{2nd}/L=0.547296$ is fitted by using the Ansatz~(\ref{chiansatz2}).
In the case of $\chi$ at $Z_a/Z_p=0.11911$, we see that the results 
differ only by little between the standard and improved version of 
$\chi$, which is due to the fact that $\lambda=18.5$ and $20$ are close to 
$\lambda^*$. 
Here we get an acceptable $p$-value for
$L_{min} \ge 8$.

Our final estimate 
\begin{equation}
\label{etafinalO4}
\eta = 0.03624(8)  \;
\end{equation}
and
the associate error bar are chosen such that the estimates of $\eta$ 
and their error bars obtained by fitting $\chi$ at $\xi_{2nd}/L=0.547296$
are covered up to $L_{min}=16$. 
The estimates obtained from $\chi$ at $Z_a/Z_p=0.11911$ are contained from 
$L_{min}=11$ up to $22$. 

\subsection{The slope of dimensionless quantities and the exponent $\nu$}
We have analyzed the slopes of dimensionless quantities at 
$Z_a/Z_p=0.11911$ or $\xi_{2nd}/L=0.547296$. We used Ans\"atze of the type
\begin{equation}
\label{nuansatz}
 S_R = a L^{y_t} \; \left(1 + \sum_i c_i L^{-\epsilon_i}  
   \right) + b L^{-\omega} \; \;, 
\end{equation}
where $y_t = 1/\nu$. The term $b L^{-\omega}$ is due to the fact that 
the scaling field of the leading correction depends on $\beta$. The derivative
of this scaling field with respect to $\beta$ in general does not 
vanish at $\lambda^*$. For a discussion see for example section III
of ref. \cite{myClock}.
We ignore leading corrections to scaling that multiply $a L^{y_t}$, since
we consider good approximations of $\lambda^*$ or we consider improved 
quantities, where leading corrections are suppressed for any $\lambda$. 

\begin{figure}
\begin{center}
\includegraphics[width=14.5cm]{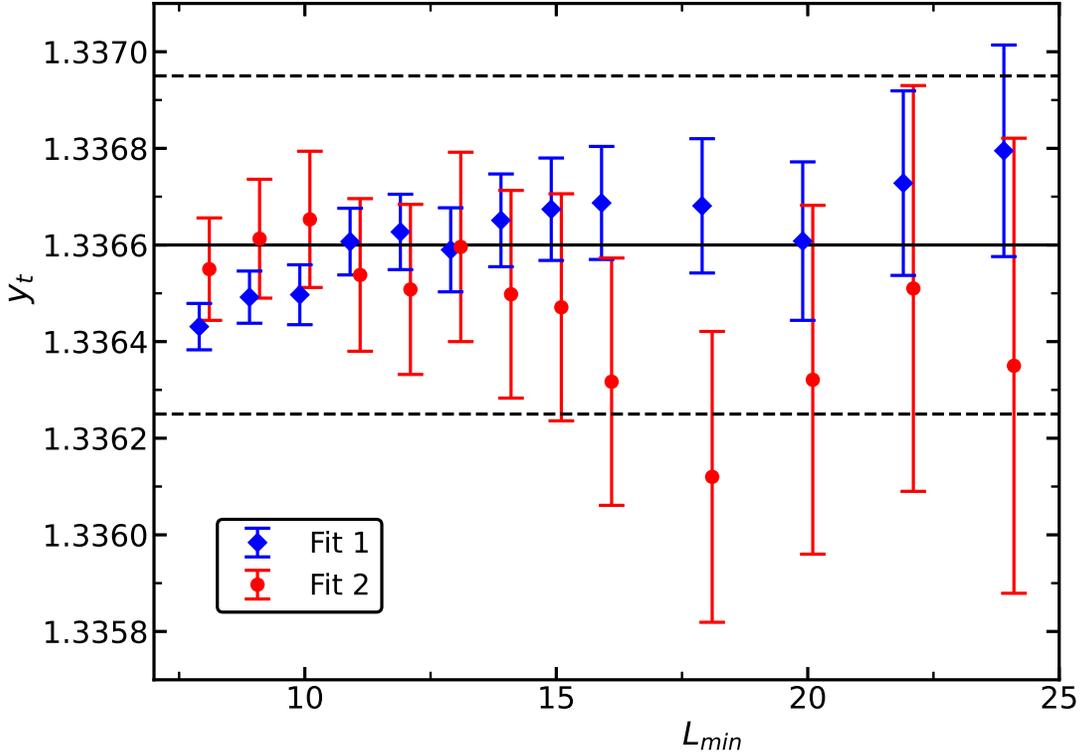}
\caption{\label{nu_ZAZP_O4}
Results for the RG-exponent $y_t$
of joint fits for $\lambda=18.5$ and $20$ for $N=4$.  
The slope of $Z_a/Z_p$ at $Z_a/Z_p=0.11911$ is fitted  
by using the Ansatz~(\ref{nuansatz})  as discussed in the text. We take
either one or two correction terms into account. In the legend, these two 
choices are referred to by fit 1 or 2.
Note that the values on the $x$-axis are slightly shifted to reduce overlap
of the symbols.
The solid line gives our final estimate
and the dashed lines indicate the error.
}
\end{center}
\end{figure}

Here we focus on the slopes of $Z_a/Z_p$ and $\xi_{2nd}/L$, since their
relative statistical error is smaller than that of the Binder cumulants $U_4$
and $U_6$.
Let us first discuss the slope of $Z_a/Z_p$ at $Z_a/Z_p=0.11911$. Here 
we expect only subleading corrections with a correction exponent close to two
due to the breaking of the rotational
invariance $\propto L^{-\omega_{NR}} $ and the additive term $b L^{-\omega}$. 
We performed fits by using an Ansatz containing only 
a correction $\propto L^{-\omega_{NR}}$
and with an Ansatz containing the term $b L^{-\omega}$ in addition. 
The results obtained for $y_t$ by using these two Ans\"atze, jointly fitting 
the data for $\lambda=18.5$ and $20$, are plotted
in Fig. \ref{nu_ZAZP_O4}. We get an acceptable $p$-value starting from 
$L_{min}=11$ for both Ans\"atze.

In order to check the effect of leading corrections to scaling on the estimate
of $y_t$, we perform separate fits for $\lambda=4.0$, $12.5$ and $\infty$. 
Fitting, taking into account only the correction 
$\propto L^{-\omega_{NR}}$,
for $L_{min}=16$, we get $y_t=1.33979(26)$, 1.33680(24), and 
1.33574(21) for $\lambda=4.0$, $12.5$ and $\infty$, respectively. We conclude 
that, given the small amplitude of leading corrections to scaling at 
$\lambda=18.5$ and $20$, we can neglect the effect of leading corrections 
to scaling in our final estimate of $y_t$, which is based on the data for 
$\lambda=18.5$ and $20$.

Next we study the slope of $\xi_{2nd}/L$ at $\xi_{2nd}/L=0.547296$. Here 
we study, also having in mind the analysis of the data for larger values 
of $N$, similar to the analysis of the magnetic susceptibility, 
improved versions of the slope. Similar to eq.~(\ref{chiimp}), we 
multiply by a power of $U_4$:
\begin{equation}
\label{Simp1}
 \bar{S}_{\xi_{2nd}/L,imp} = \bar{U}_4^x \bar{S}_{\xi_{2nd}/L} \;.
\end{equation}
Furthermore, one might construct an improved slope by combining the slope of
$\xi_{2nd}/L$ with that of the Binder cumulant $U_4$:
\begin{equation}
\label{Simp2}
 \bar{S}_{mix} =  \bar{S}_{\xi_{2nd}/L} + x \bar{S}_{U_4} \;.
\end{equation}
Similar to the analysis of the dimensionless quantities, we performed
joint fits for two sets of $\lambda$ values.
We consider $\lambda=\infty$, $20$, $18.5$, $12.5$, and $4$. As check,
we take a second set, where $\lambda=2$ is added. Also here we used 
two different types of fits. In the first fit, we used a single correction 
with $\epsilon_1=2$.  The additive correction with the exponent $\omega$ 
is neglected. In the second fit, this correction is present. For both
corrections we use eq.~(\ref{parac1}) as parameterization of the coefficient.
Given the present statistical error of the data, we can not resolve a larger
number of corrections with $\epsilon \approx 2$.  

\begin{figure}
\begin{center}
\includegraphics[width=14.5cm]{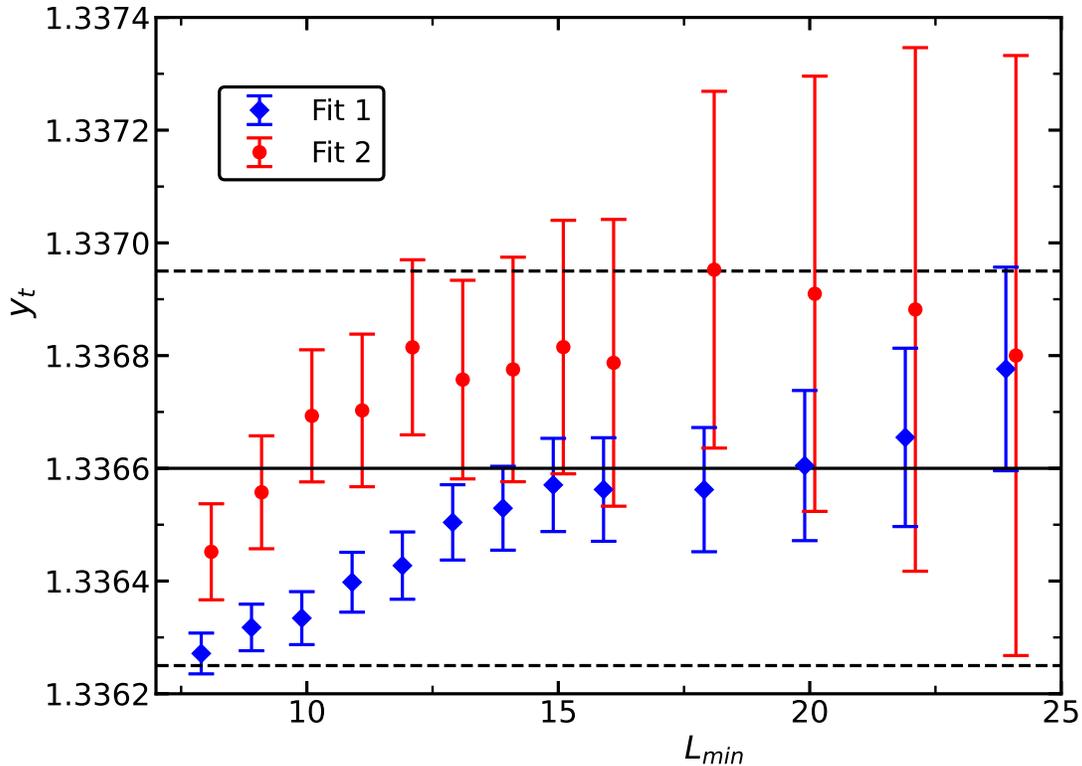}
\caption{\label{nu_xi2L_O4}
Results of joint fits for 
$\lambda=\infty$, $20$, $18.5$, $12.5$, $4$ and $2$ 
 for $N=4$.
The improved slope of $\xi_{2nd}/L$ at $\xi_{2nd}/L=0.547296$, 
eq.~(\ref{Simp1}), is fitted by using
the Ansatz~(\ref{nuansatz}) as discussed in the text. We take
either one or two correction terms into account. In the legend, these two
choices are referred to by fit 1 or 2.
Note that the values on the $x$-axis are slightly shifted to reduce overlap
of the symbols.
The solid line and the dashed lines
give the result and the error bar obtained above from the slope of 
$Z_a/Z_p$.
}
\end{center}
\end{figure}

Analyzing the improved slope~(\ref{Simp1}) we find $x = 0.25(15)$ taking into
account the two types of fits that we performed. Note that 
$\chi^2/$DOF$=1.20$ is reached for $L_{min}=16$ in the case of fit 1 and
$\chi^2/$DOF$=1.10$ for $L_{min}=10$ in the case of fit 2. In both cases 
all values of $\lambda$ are taken into account. The estimates of $y_t$ 
obtained by performing these fits are shown in Fig. \ref{nu_xi2L_O4}. We 
find that the result is fully consistent with that obtained above for
$Z_a/Z_p$.

Analyzing the  mixed slope~(\ref{Simp2}) we find $x = -0.05(5)$. The results
obtained for $y_t$ are very similar to those for the improved 
slope~(\ref{Simp1}).

As our final result we quote the one obtained from the slope of $Z_a/Z_p$ at
$\lambda=18.5$ and $20$
\begin{equation}
\label{ytN4}
 y_t = 1.33660(35) \;,
\end{equation}
corresponding to $\nu=0.74817(20)$.

\section{The simulations and the analysis of the data for $N=5$}
\label{ana5}
We have simulated at $\lambda=\infty$, $10$, $5$, and $1$. 
In the case of $\lambda=\infty$ we simulated the lattice sizes $L=8$, $9$,
$10$, ..., $16$, $18$, $20$, ..., $24$, $28$, $32$, ..., $40$, $48$, $56$, 
..., $80$, $100$, and $200$.  The number of measurements is decreasing 
with increasing lattice size $L$. For example, we performed 
$3.4 \times 10^9$, $1.2 \times 10^8$, and $ 8.5 \times 10^7$ measurements
for $L=10$, $100$, and $200$, respectively. These simulations took about 
$6.4$ years of CPU time. 

For $\lambda=10$ and $5$  we simulated the same set of lattice size as 
for $\lambda=\infty$ but with a maximal lattice size $L=72$.  The number of
measurements for each lattice size is a bit smaller than for 
$\lambda=\infty$. These simulations took about
$3.3$ and $2.8$ years of CPU time  
for $\lambda=10$ and $5$, respectively.

For $\lambda=1$, we simulated the lattice sizes $L=12$, $16$, $20$, ..., 
$48$, $56$, $64$, $72$, and $80$. For example for $L=12$ and $80$, we performed
$2.3 \times 10^9$ and   $2.4 \times 10^8$ measurements, respectively. 
These simulations took about $6.9$ years of CPU time.

\subsection{The dimensionless quantities}
We performed fits by using the same Ans\"atze as for $N=4$. First we
analyzed the data for $\lambda=\infty$, $10$, and $5$ jointly.   As check, 
we have added in a second set of fits the data for $\lambda=1$.
In the case of the first set, using the Ansatz~(\ref{masterdimless}) 
with $k_{max}=1$ and the parameterization~(\ref{parac0}),
a reasonable  goodness
of the fit is reached at $L_{min}=22$ with $\chi^2/$DOF$=1.15$ corresponding
to $p=0.127$. Using Ansatz~(\ref{masterdimless}) 
with $k_{max}=2$ and the parameterization~(\ref{parac1}), we get for example 
$\chi^2/$DOF$=1.18$ corresponding to $p=0.039$ for $L_{min}=12$ and
$\chi^2/$DOF$=0.99$ corresponding to $p=0.515$ for $L_{min}=16$. Adding the 
data for $\lambda=1$, we get $\chi^2/$DOF$=1.07$ 
corresponding to $p=0.244$ for $L_{min}=18$. Using the 
Ansatz~(\ref{masterdimless})
with $k_{max}=3$ and the parameterization~(\ref{parac2}), we get
$\chi^2/$DOF$=1.14$ corresponding to $p=0.080$ for $L_{min}=14$ and 
$\chi^2/$DOF$=1.00$ corresponding to $p=0.470$ for $L_{min}=18$.
Below, we refer to these four choices of the data set and the Ansatz 
as fit $1$, $2$, $3$, and $4$, respectively.

\begin{figure}
\begin{center}
\includegraphics[width=14.5cm]{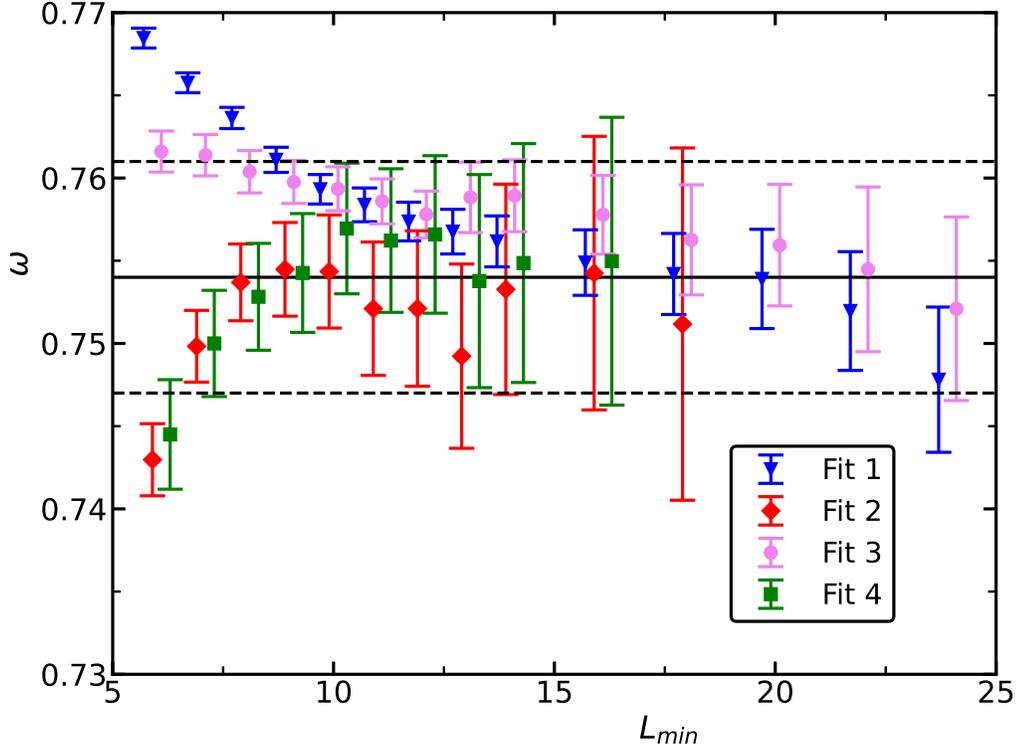}
\caption{\label{Omega_O5}
Numerical estimates of the correction exponent $\omega$ for $N=5$ obtained
from the fits discussed in the text. Fit 1 and 2 are based on the data 
for $\lambda=\infty$, $10$ and $5$. In the case of fits 3 and $4$, 
in addition, $\lambda=1$ is taken into account. In the case of fit 1, we 
use $k_{max}=1$ and the  parameterization~(\ref{parac0}). In the case of fits 
2 and 3 we use 
$k_{max}=2$ and the  parameterization~(\ref{parac1}), while for fit 4
$k_{max}=3$ and the  parameterization~(\ref{parac2}) is taken.
Note that the values on the $x$-axis are slightly shifted to reduce overlap
of the symbols. The solid line and the dashed lines
give the final result and the error bar.
}
\end{center}
\end{figure}

In Fig. \ref{Omega_O5} we plot the estimates of the correction exponent 
$\omega$ obtained by fitting as discussed above.
As our final result we quote
\begin{equation}
\omega=0.754(7) \;.
\end{equation}
This estimate covers 
fit 1
 for $L_{min}=11$ up to $22$, 
fit 2
 for $L_{min}=7$ up to $12$, 
fit 3
for $L_{min}=9$ up to $22$, 
fit 4
for $L_{min}=8$ up to $11$ and $13$.

We have determined the fixed point values of the dimensionless quantities
in a similar fashion as for $N=4$. 
We skip a detailed discussion of the analysis.
Our results are summarized in table \ref{Rfinal}.

Next let us discuss the amplitude of leading corrections.
For example for  fit  4 with  $L_{min}=14$, we get
$b=-0.00088(21)$, $-0.00870(30)$, $-0.01502(43)$, and $-0.0407(13)$ for 
$\lambda=\infty$, $10$, $5$ and $1$, respectively. Varying the form 
the Ansatz~(\ref{masterdimless})
we find $b<0$ throughout for $\lambda=\infty$.  Assuming that 
$b$ is a monotonous function of $\lambda$, this implies that for $N=5$
no $\lambda^*$ exists.  However the amplitude of $b$ at $\lambda=\infty$ is
rather small. Therefore in the following analysis of the data, we can regard
$\lambda=\infty$ as a reasonable approximation of  $\lambda^*$. 

With increasing $N$ the value of $(Z_a/Z_p)^*$ approaches $0$, while 
$(\xi_{2nd}/L)^*$ approaches a finite value. Therefore, going to larger 
values of $N$, we focus on $\xi_{2nd}/L$ instead of $Z_a/Z_p$. In particular
in the Ansatz~(\ref{masterdimless}) we set $a_{\xi_{2nd}/L}=1$ instead of 
$a_{Z_a/Z_p}=1$. This way the correction amplitude $b(\lambda)$ for 
different values of $N$ can be compared more easily. Estimates of 
$b(\lambda)$, setting $a_{\xi_{2nd}/L}=1$ for $N\ge5$ are given in 
table \ref{corrections_b}.

Our estimates of the inverse critical temperature $\beta_c$
are summarized in table \ref{betac5to12}.

\subsection{The magnetic susceptibility and the exponent $\eta$}
\begin{figure}
\begin{center}
\includegraphics[width=14.5cm]{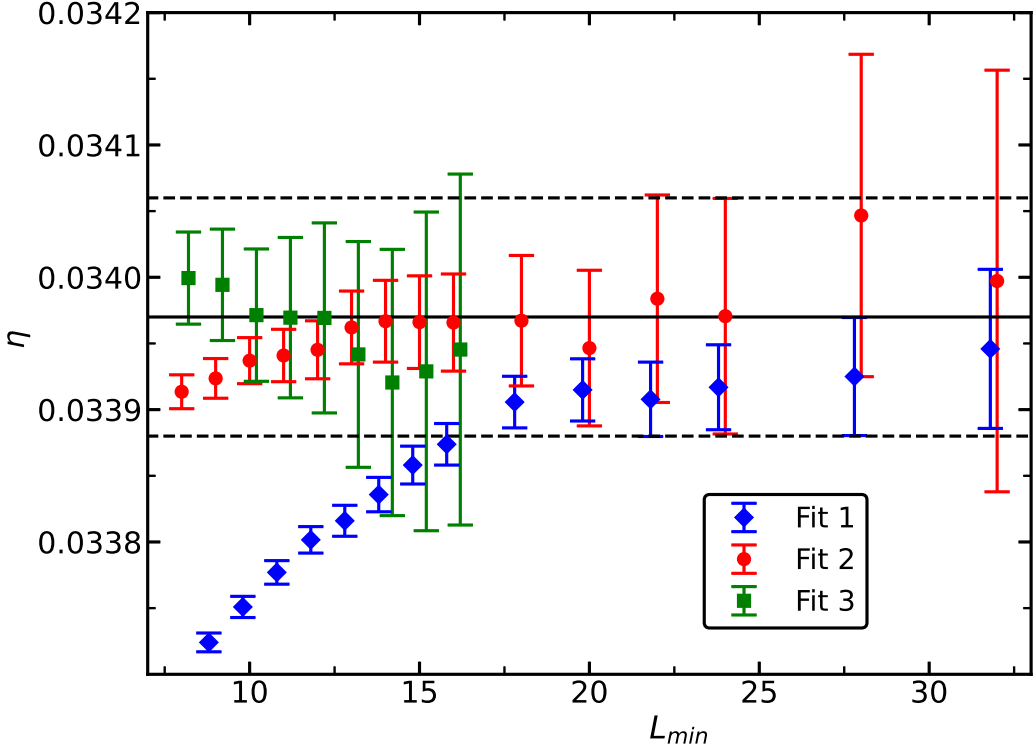}
\caption{\label{Chi_eta_O5}
Estimates of $\eta$  
from joint fits for $\lambda=\infty$, $10$, $5$, and $1$ for $N=5$.  
The data for $\chi_{imp}$ at $\xi_{2nd}/L=0.53691$ are fitted 
by using the Ans\"atze~(\ref{chiansatz0},\ref{chiansatz1},\ref{chiansatz2}).
In the legend, these are referred to as fit 1, 2, and 3.
Note that the values on the $x$-axis are slightly shifted to reduce overlap
of the symbols.
The solid line gives our final estimate
and the dashed lines indicate the error.   
}
\end{center}
\end{figure}

First we analyzed the improved
magnetic susceptibility, eq.~(\ref{chiimp}), at $\xi_{2nd}/L=0.53691$.
Here we used the Ansatz
\begin{equation}
\label{chiansatz0}
 \bar{\chi}_{imp} = a L^{2-\eta} + b \;,
\end{equation}
where $a$ and $b$ are free parameters for each value of $\lambda$ and 
in addition the Ans\"atze~(\ref{chiansatz1},\ref{chiansatz2}).
We included data for all values of $\lambda$ that we have simulated.
In Fig. \ref{Chi_eta_O5} our results for $\eta$ are plotted versus the 
minimal lattice size $L_{min}$ that is taken into account. We get 
acceptable $p$-values starting from $L_{min}=16$, $8$ and $8$ for the 
Ans\"atze~(\ref{chiansatz0},\ref{chiansatz1},\ref{chiansatz2}), respectively.
As our final result we take

\begin{equation}
\label{finaleta5}
 \eta=0.03397(9)  \;.
\end{equation}
Furthermore, we get $x=0.075(25)$.  Fixing $Z_a/Z_p=0.07263$  
instead of $\xi_{2nd}/L=0.53691$ we get $\eta=0.03398(11)$ and 
$x=-2.06(6)$.  As a check, we analyzed the data for $\chi$ and $\chi_{imp}$
with $x=0.075$ at $\xi_{2nd}/L=0.53691$ for $\lambda=\infty$ 
separately.  We find that the results for $\eta$ obtained by 
fitting $\chi$ and $\chi_{imp}$ differ only by a small fraction of the 
error bar. 
We find that the results for $\eta$
are fully consistent with eq.~(\ref{finaleta5})
that we regard as our final estimate. 

\subsection{The slope of dimensionless quantities and the critical exponent $\nu$}
\begin{figure}
\begin{center}
\includegraphics[width=14.5cm]{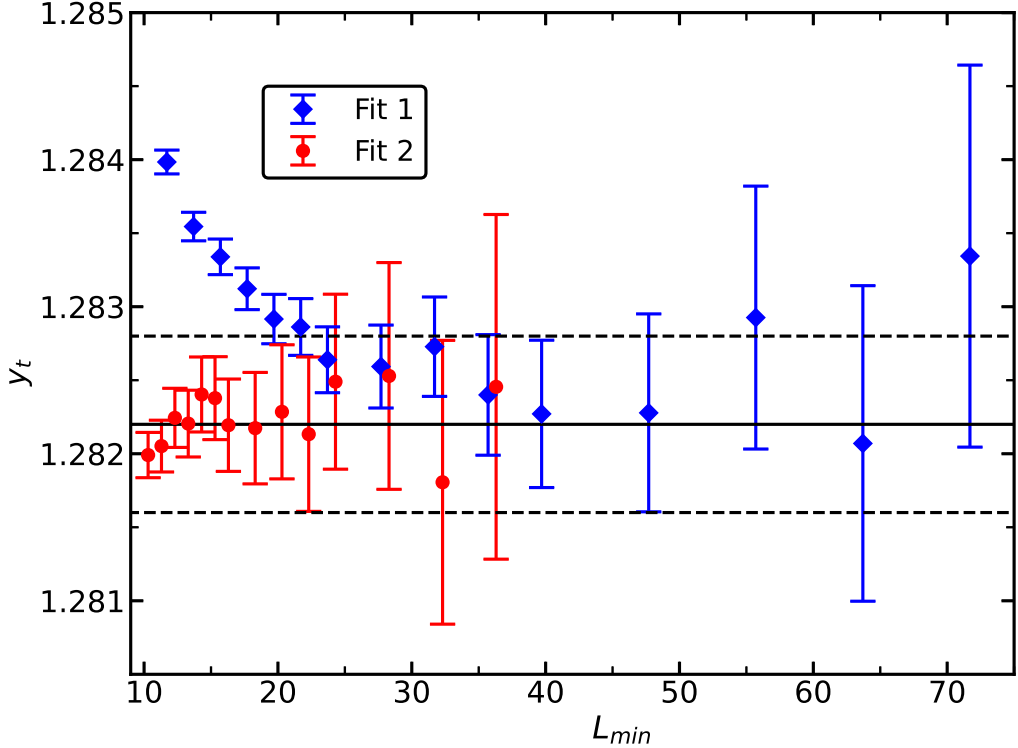}
\caption{\label{yt5ZaZp}
Estimates of $y_t$
from  fits  of the slope of  $Z_a/Z_p$ at $Z_a/Z_p=0.07263$
for $\lambda=\infty$ and $N=5$ plotted versus the minimal lattice size
$L_{min}$ that is taken into account.
In the legend, fit 1 refers to an Ansatz without correction term, while
fit 2 refers to an Ansatz with a correction proportional to $L^{-\omega_{NR}}$. 
Note that the values on the $x$-axis are slightly shifted to reduce overlap
of the symbols.
The solid line gives our preliminary estimate
and the dashed lines indicate the error: $y_t=1.2822(6)$.
}
\end{center}
\end{figure}

First we have analyzed the slope of the ratio of partition functions $Z_a/Z_p$ 
at $Z_a/Z_p=0.07263$.  Here we expect subleading corrections proportional to 
$L^{-\omega_{NR}}$.  The corrections due to the additive 
contribution $b L^{-\omega}$ effectively corresponds to a correction with the
exponent $y_t + \omega$. Putting in the numerical values $1.282+0.754(7) \approx 2.036$, 
where we anticipate our estimate of $y_t$ given below, we get for $N=5$ a 
value close to $\omega_{NR}$.
Since there is little chance to disentangle these two different corrections
in the fit, we use an Ansatz containing a single correction term.
The results for the RG-exponent $y_t$ obtained from the data 
for $\lambda=\infty$ are plotted in Fig. \ref{yt5ZaZp}.  
Acceptable $p$-values are obtained for $L_{min} \ge 12$. 
For the Ansatz without any correction,
acceptable $p$-values are obtained for $L_{min} \ge 24$. As estimate we
take $y_t=1.2822(6)$. To get an idea on the effect of the leading correction
to scaling, we quote the results obtained for $L_{min}=12$ and the Ansatz 
containing a correction term proportional to $L^{-\omega_{NR}}$: 
$y_t = 1.28224(20)$, $1.28424(27)$, $1.28679(30)$, and $1.29557(24)$ for
$\lambda=\infty$, $10$, $5$, and $1$, respectively.

\begin{figure}
\begin{center}
\includegraphics[width=14.5cm]{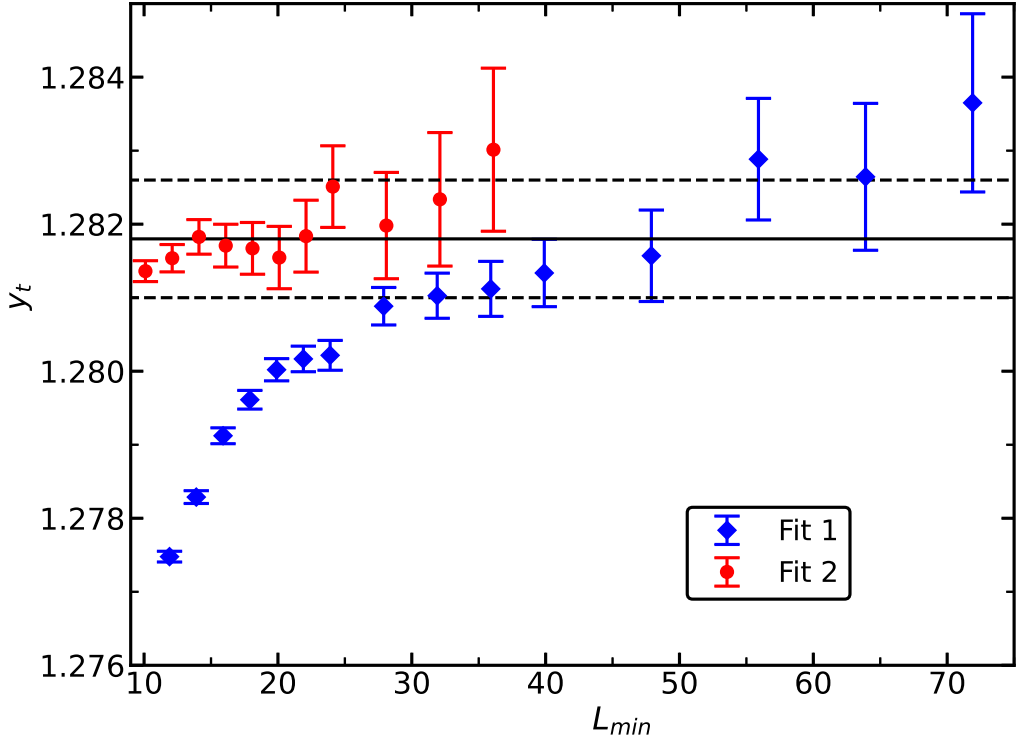}
\caption{\label{yt5xi2L}
Estimates of $y_t$
from  fits  of the slope of  $\xi_{2nd}/L$ at $\xi_{2nd}/L=0.53691$
for $\lambda=\infty$ and $N=5$ plotted versus the minimal lattice size
$L_{min}$ that is taken into account.
In the legend, fit 1 refers to an Ansatz without correction term, while
fit 2 refers to an Ansatz with a correction proportional to $L^{-2+\eta}$.
Note that the values on the $x$-axis are slightly shifted to reduce overlap
of the symbols.
The solid line gives our preliminary estimate
and the dashed lines indicate the error: $y_t=1.2818(8)$.
}
\end{center}
\end{figure}

Next we analyze the slope of $\xi_{2nd}/L$ at $\xi_{2nd}/L=0.53691$.
Here we expect in addition subleading corrections with the exponents $2-\eta$ and $2$.  Since it is virtually impossible to disentangle the subleading 
corrections here, we performed fits without any corrections and with an Ansatz
containing a single correction term proportional to $L^{-2+\eta}$. 
Our results for $\lambda=\infty$ are given in Fig. \ref{yt5xi2L}.
Without any correction, we get a acceptable $p$-values for $L_{min} \ge 28$ and
including one correction term, we get acceptable $p$-values for $L_{min} \ge 8$.

For the fit with a correction term and $L_{min}=16$ we get 
$y_t=1.28171(29)$, $1.28134(33)$, 
$1.28305(44)$, and $1.28501(35)$ for $\lambda=\infty$, $10$, $5$, and $1$,
respectively. We see that the estimate of $y_t$ is less effected by leading 
corrections to scaling than in the case of the slope of $Z_a/Z_p$ at 
$Z_a/Z_p=0.07263$.

As our final estimate we quote 
\begin{equation}
\label{finalytO5}
y_t=1.2818(10)
\end{equation}
that covers both preliminary estimates and takes into account that 
the estimate obtained from the slope of $Z_a/Z_p$ might be slightly 
overestimated due to leading corrections to scaling.

\begin{figure}
\begin{center}
\includegraphics[width=14.5cm]{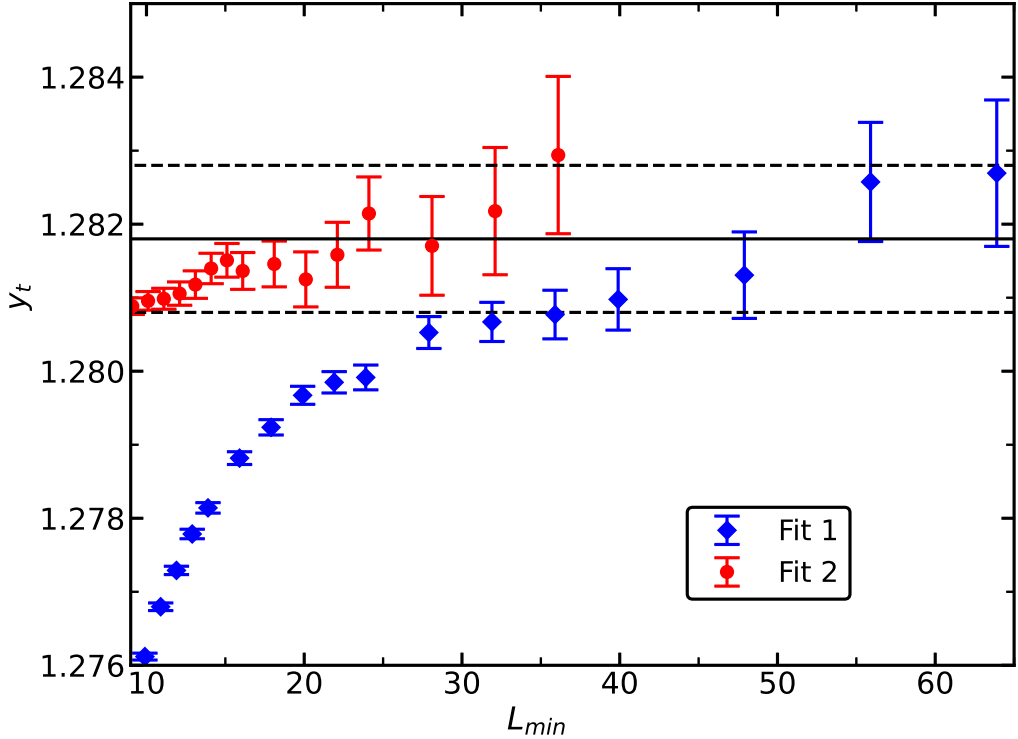}
\caption{\label{yt5xi2}
Estimates of $y_t$ from joint fits of the improved slope, eq.~(\ref{Simp1}),
of  $\xi_{2nd}/L$ at $\xi_{2nd}/L=0.53691$ using the data for
$\lambda=\infty$, $10$, $5$, and $1$ for $N=5$.
In the legend, fit 1 refers to an Ansatz without correction term, while
fit 2 refers to an Ansatz with a correction proportional to $L^{-2}$.
Note that the values on the $x$-axis are slightly shifted to reduce overlap
of the symbols.
The solid line gives our final estimate, eq.~(\ref{finalytO5}),
and the dashed lines indicate the error. 
}
\end{center}
\end{figure}

As check we analyze the improved slopes, eqs.~(\ref{Simp1},\ref{Simp2}),
of $\xi_{2nd}/L$ at $\xi_{2nd}/L=0.53691$.  First we have analyzed 
the improved slope, eq.~(\ref{Simp1}).  We used an Ansatz without correction
term and one with a correction proportional to $L^{-2}$. Note that replacing 
the correction exponent $2$ by $2-\eta$ or $\omega_{NR}$ changes the estimate 
of $y_t$ only by little.
An acceptable goodness                                                    
of the fits is reach for $L_{min}=28$ and $14$, respectively.
The estimates of $y_t$ are given in 
Fig.~(\ref{yt5xi2}). These estimates are consistent with our final 
estimate, eq.~(\ref{finalytO5}). As estimate of the exponent in 
eq.~(\ref{Simp1}) we get $x=1.05(20)$. 

We obtain qualitatively similar result for the other improved 
quantity, eq.~(\ref{Simp2}).

\section{The simulations and the analysis of the data for $N \ge 6$}
\label{ana6plus}
In addition to $N=4$ and $5$, we have simulated the $\phi^4$ model for
$N=6$, $8$, $10$, and $12$. In all cases, we simulated at $\lambda=\infty$, 
$10$, and $5$.  For $N=10$, in addition, $\lambda=1$ is considered.
The largest lattice size that we simulate for $\lambda=\infty$ is 
$L=200$, $100$, $100$, and $72$ for $N=6$, $8$, $10$, and $12$, respectively.
The statistics for a given lattice size is similar to that of the simulations
for $N=4$ and $5$ discussed above. In total we used 
$20.5$, $13.3$, $24.2$, and $12.9$ years of CPU time for the 
simulations for $N=6$, $8$, $10$, and $12$, respectively. 

\subsection{The dimensionless quantities}
For $N \ge 6$, we analyzed dimensionless quantities in a similar way 
as for $N=4$ and $5$. We fitted our data by using the 
Ansatz~(\ref{masterdimless}). 
Here we use a parameterization, where $a_{\xi_{2nd/L}}=1$ and 
$a_{Z_a/Z_p}$ is a free parameter.
Taking into  account data for $\lambda=\infty$,
$10$ and $5$, we consider either $k_{max}=1$ or $2$ and either the 
parameterization~(\ref{parac0}) or (\ref{parac1}). In the case of $N=10$, 
taking into account the data for $\lambda=1$, we also used 
$k_{max}=3$ and the parameterization~(\ref{parac2}).

Throughout, the $\chi^2/$DOF and the corresponding goodness of the 
fit as a function of $L_{min}$ behave similar to that discuss for 
$N=4$ and $5$. Therefore we abstain from a detailed discussion.

Let us first discuss the amplitude of corrections to scaling. In 
table \ref{corrections_b} we give the amplitude $b(\lambda)$ obtained
from the fit with $k_{max}=2$ and the parameterization~(\ref{parac1})
for $L_{min}=12$.  Here we abstain from estimating the systematic error
of $b(\lambda)$, since we are mainly interested in the qualitative
picture. For $N \ge 6$, $b(\infty)>0$, as it is the case of finite $\lambda$.
The value of $b(\lambda)$ increases with decreasing $\lambda$.  This
means that there is no $\lambda^*$ and the amplitude of leading corrections 
to scaling is minimal for $\lambda=\infty$.  Furthermore we note that
the values of  $b(\lambda)$ for a given value of $\lambda$ are similar 
for $N=8$, $10$, and $12$. 

The fixed point values of dimensionless quantities are determined in the 
same fashion as for $N=4$ and $5$. Our final results are summarized in 
table \ref{Rfinal}.
We get $\omega=0.758(14)$ and $0.816(16)$ for $N=6$ and $10$, respectively. 
In the 
other cases, it is hard to give a reasonable estimate of the error due to 
a lack of statistics or data for $\lambda=1$.
Our results for the inverse critical temperature $\beta_c$ are summarized in 
table \ref{betac5to12}. To our knowledge, for $\lambda=\infty$,
the most accurate results given in the literature for $N=1$, $2$, and $3$
are $\beta_c=0.221654628(2)$, ref. \cite{Landau18}, $0.45416466(10)$, 
ref. \cite{Xu19}, and $0.693003(2)$, ref. \cite{Deng05}, respectively.

\begin{table}
\caption{\sl \label{Rfinal}
Estimates of the fixed point values $R^*$ of the dimensionless
quantities that we have analyzed. For completeness, we have copied 
the values for $N=4$ from 
eqs.~(\ref{ZaZpstarO4},\ref{xi2LstarO4},\ref{U4starO4},\ref{U6starO4}).
}
\begin{center}
\begin{tabular}{rlllll}
 $N$ &  \mc{1}{c}{$(Z_a/Z_p)^*$}  &  \mc{1}{c}{$(\xi_{2nd}/L)^*$} &  
 \mc{1}{c}{$U_4^*$} &  \mc{1}{c}{$U_6^*$} \\
\hline
  4  &  0.11911(2)  & 0.547296(26) & 1.094016(12) & 1.281633(33)    \\
  5  &  0.07263(4) &  0.53691(7)  & 1.069735(25) & 1.20860(8) \\
  6  &  0.04401(4) &  0.53038(6)  & 1.054960(25) & 1.16439(8) \\
  8  &  0.015835(35)& 0.5232(1)   & 1.03825(3) & 1.11445(10) \\
 10  &  0.005610(8) & 0.51967(10) & 1.02924(2) & 1.08753(6)  \\
 12  &  0.00196(1)  & 0.5178(2)   & 1.02360(4) & 1.07065(10) \\
\hline
\end{tabular}
\end{center}
\end{table}

\begin{table}
\caption{\sl \label{corrections_b}
Estimates of the leading correction amplitude $b(\lambda)$ obtained
by fitting with the Ansatz~(\ref{masterdimless}), setting $a_{\xi_{2nd}/L}=1$. 
$k_{max}$ and param. refer to the precise form of the Ansatz.
Throughout the 
minimal linear lattice size that is taken into account is set to
$L_{min}=12$. The number in parenthesis gives the statistical error.
}
\begin{center}
\begin{tabular}{rccllll}
\hline
\mc{1}{c}{$N$} &$k_{max}$ & param. & \mc{1}{c}{$b(\infty)$}&   
\mc{1}{c}{$b(10)$}  &
\mc{1}{c}{$b(5)$}   &    \mc{1}{c}{$b(1)$} \\
\hline
 5&$2$ & eq.~(\ref{parac1})& 0.00167(34)& 0.01844(34)&0.03192(36)&   \\
 5&$3$ & eq.~(\ref{parac2})& 0.00182(32)& 0.0188(4)  &0.0325(6)  &0.0885(19)\\
 6&$2$ & eq.~(\ref{parac1})& 0.0090(4)  & 0.0283(6)  &0.0426(8)  &   \\
 8&$2$ & eq.~(\ref{parac1})& 0.0180(6)  & 0.0430(10) &0.0591(14) &   \\
10&$2$ & eq.~(\ref{parac1})& 0.0237(8)  & 0.0557(17) &0.0743(23) &   \\
10&$3$ & eq.~(\ref{parac2})& 0.0232(6)  & 0.0522(12) &0.0687(16) &0.1194(31) \\ 
12&$2$ & eq.~(\ref{parac1})& 0.0245(19) & 0.0509(47) &0.0641(63) &  \\
\hline
\end{tabular}
\end{center}
\end{table}

\begin{table}
\caption{\sl \label{betac5to12}
Estimates of the inverse critical temperature $\beta_c$ for
$N=5$, $6$, $8$, $10$, and $12$  for $\lambda=\infty$, $10$ and $5$.
For $N=5$ and $10$, $\lambda=1$ we get $\beta_c=0.8044989(5)$
and $1.1665100(13)$, respectively.
}
\begin{center}
\begin{tabular}{clll}
\hline
\mc{1}{c}{$N$\textbackslash$\lambda$}  &   \mc{1}{c}{$\infty$} & \mc{1}{c}{10} &  \mc{1}{c}{5}  \\
\hline
 5  &  1.1813639(5) & 1.1054374(6) & 1.0452357(8) \\
 6  &  1.4286859(9) & 1.2991764(10) & 1.2067603(8) \\
 8  &  1.926761(3)& 1.6642478(18) & 1.5040374(14) \\
10  &  2.427525(4) &  2.0039555(35)& 1.7744067(20)  \\
12  &  2.929802(11) & 2.322675(5)  & 2.023970(4) \\
\hline
\end{tabular}
\end{center}
\end{table}

\subsection{The magnetic susceptibility and the critical exponent $\eta$}
Since $(Z_a/Z_p)^*$ rapidly decreases with increasing $N$ and its relative 
error increases with  increasing $N$, for $N \ge 6$, we only consider $\chi$
at a fixed value of $\xi_{2nd}/L$. To this end, we take our estimates of 
$(\xi_{2nd}/L)^*$ summarized in table \ref{Rfinal}.

We perform fits of the improved susceptibility, eq.~(\ref{chiimp}), 
where the exponent 
$x$ is a free parameter. We use Ans\"atze of the form eq.~(\ref{chi1}).
In particular
\begin{equation}
\label{chian0}
\bar{\chi}_{imp} = a(\lambda) L^{2-\eta} + b(\lambda)  \;,
\end{equation}
\begin{equation}
\label{chian1}
\bar{\chi}_{imp} = a(\lambda) L^{2-\eta} (1+ c L^{-2}) + b(\lambda) \;,
\end{equation}
and
\begin{equation}
\label{chian2}
\bar{\chi}_{imp} = a(\lambda) L^{2-\eta} (1+ c L^{-2} +d L^{-\omega_{NR}}) + 
b(\lambda) \;,
\end{equation}
where $a(\lambda)$ and $b(\lambda)$ are free parameters for each value 
of $\lambda$, while $c$ and $d$ are take the same value for all $\lambda$.

\begin{figure}
\begin{center}
\includegraphics[width=14.5cm]{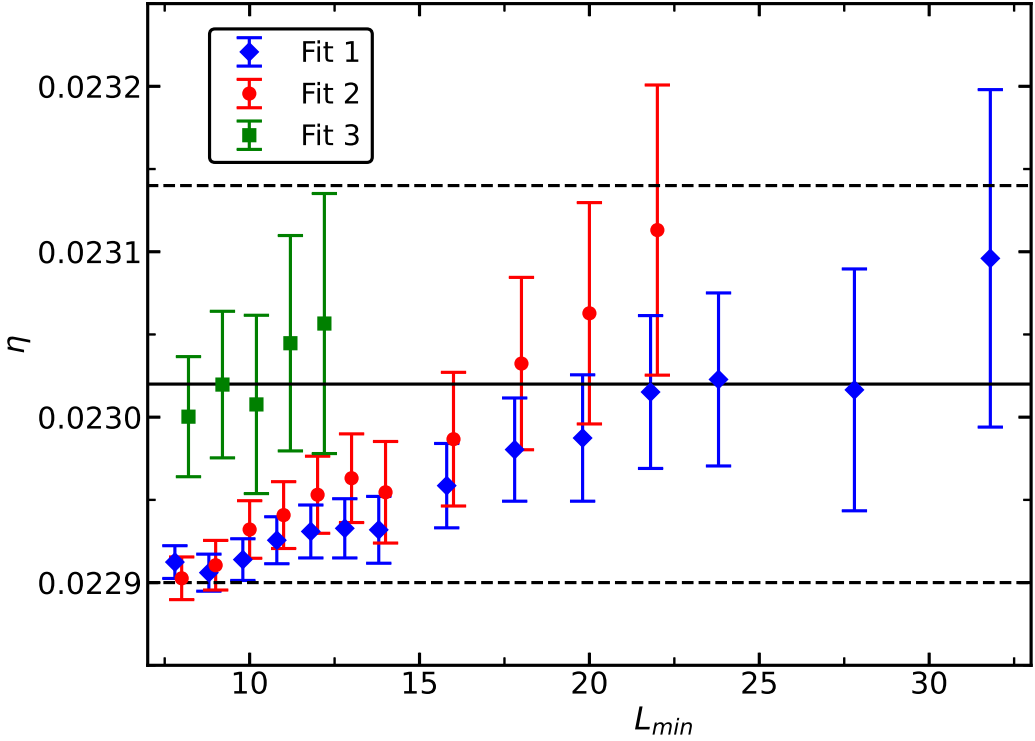}
\caption{\label{O10eta}
Estimates of $\eta$ from joint fits of the improved magnetic susceptibility, 
eq.~(\ref{chiimp}), at $\xi_{2nd}/L=0.51967$ using the data for
$\lambda=\infty$, $10$, and $5$  for $N=10$.
In the legend, fits 1, 2, and 3 refer to the Ans\"atze, 
eqs.~(\ref{chian0},\ref{chian1},\ref{chian2}), respectively.
Note that the values on the $x$-axis are slightly shifted to reduce overlap
of the symbols.
The solid line gives our final estimate,
and the dashed lines indicate the error. 
}
\end{center}
\end{figure}

As an example, in Fig. \ref{O10eta}, we give estimates of $\eta$ for $N=10$.
We get an acceptable $p$-value already for $L_{min}=8$ for all three 
Ans\"atze that we consider. The final results for the critical exponent
$\eta$ are given in table \ref{exponents}.
Results for the exponent in eq.~(\ref{chiimp}) are $x=0.14(4)$, $0.21(6)$, 
$0.30(5)$, and $0.25(5)$ for $N=6$, $8$, $10$, and $12$, respectively.

\begin{table}
\caption{\sl \label{exponents}
Estimates of $\eta$ and $y_t=1/\nu$ for $N=6$, $8$, $10$, and $12$.
For a discussion see the text.
}
\begin{center}
\begin{tabular}{rll}
\hline
\mc{1}{c}{$N$} & \mc{1}{c}{$\eta$} & \mc{1}{c}{$y_t$}  \\
\hline
 6    & 0.03157(14) & 1.2375(9)  \\
 8    & 0.02675(15) & 1.1752(10) \\
10    & 0.02302(12) & 1.1368(12) \\
12    & 0.0199(3)   & 1.1108(17) \\
\hline
\end{tabular}
\end{center}
\end{table}

\subsection{The slope of dimensionless quantities and the critical exponent $\nu$}
We analyzed the improved slope of $\xi_{2nd}/L$,  eq.~(\ref{Simp1}), 
at a fixed value of $\xi_{2nd}/L$.
Here we consider the Ans\"atze 
\begin{equation}
\label{yt0AN}
\bar{S}_{imp} = a(\lambda) L^{y_t}
\end{equation}
and
\begin{equation}
\label{yt1AN}
\bar{S}_{imp} = a(\lambda) L^{y_t} + b(\lambda) L^{-\omega} \;,
\end{equation}
where $a(\lambda)$ and $b(\lambda)$ are free parameters for each
value of $\lambda$. While $y_t$ is a free parameter, we fix $\omega$.
To this end, we use the values obtained by the biased Pad\'e approximation 
discussed below. We checked that varying the value of $\omega$ within 
plausible errors, the estimates of $y_t$ change only by little.

\begin{figure}
\begin{center}
\includegraphics[width=14.5cm]{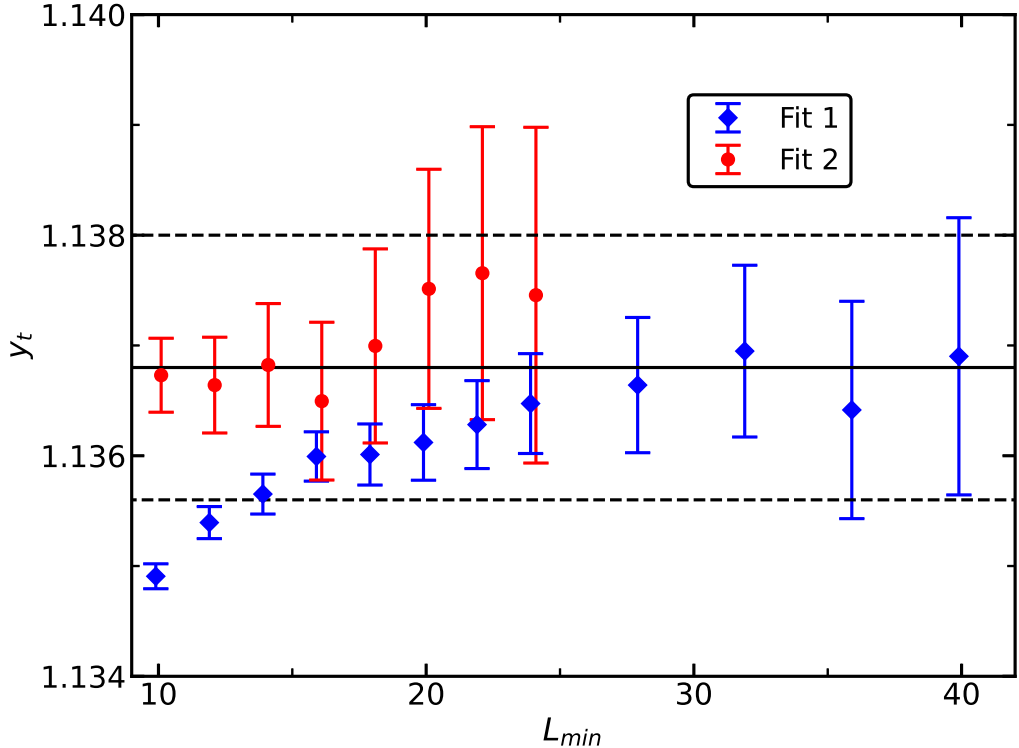}
\caption{\label{O10yt}
Estimates of $y_t$ from joint fits of the improved slope of $\xi_{2nd}/L$,
eq.~(\ref{Simp1}), at $\xi_{2nd}/L=0.51967$ using the data for
$\lambda=\infty$, $10$, and $5$  for $N=10$.
In the legend, fits 1 and 2 refer to the Ans\"atze,
eqs.~(\ref{yt0AN},\ref{yt1AN}), respectively.
Note that the values on the $x$-axis are slightly shifted to reduce overlap
of the symbols.
The solid line gives our final estimate
and the dashed lines indicate the error.
}
\end{center}
\end{figure}

As an example, in Fig. \ref{O10yt} we give estimates of $y_t$ for $N=10$.
The final results for the RG-exponent
$y_t$ are given in table \ref{exponents}. Here we get an acceptable $p$-value
for $L_{min} \ge 16$ and $8$ for the Ans\"atze~(\ref{yt0AN},\ref{yt1AN}), 
respectively.
The estimates for the exponent in eq.~(\ref{Simp1}) are $x=1.7(3)$, $3.7(6)$,
$5.7(5)$, and $7(1)$ for $N=6$, $8$, $10$, and $12$, respectively.

\section{Summary and comparison with other results}
\label{comparison}
We have studied the $O(N)$-symmetric $\phi^4$ model on the 
simple cubic lattice by using Monte Carlo simulations in conjunction
with a finite size scaling (FSS) analysis. 
In the cases $N=1$, $2$, $3$ and $4$ it had been demonstrated before
that there is a value $\lambda^*$ of the parameter  $\lambda$, where 
leading corrections to scaling vanish. In the large $N$ limit such a
$\lambda^*$ does not exist \cite{CaPe99}. Here we confirm the existence 
of $\lambda^*$ for $N=4$ and provide a more accurate numerical estimate.
In contrast, for $N=5$ it is quite clear from the data that no $\lambda^*$
exists. In the limiting case $\lambda=\infty$, the amplitude of the corrections
is relatively small.  Going to larger values of $N$,  there is no doubt
that there is no $\lambda^*$. The minimal amplitude of corrections to scaling
is found for the limiting case $\lambda=\infty$. However these corrections 
can not be ignored in the analysis of the data.

Estimating critical exponents, we focus on $\lambda^*$ and 
$\lambda=\infty$ for $N=4$ and $5$, respectively. For larger values of $N$
we have to deal with leading corrections in a different way. Instead
of putting them explicitly into the Ans\"atze, we use improved observables.
The advantage of this approach is that the exponent of the leading 
correction is not needed. 

The $O(N)$-symmetric $\phi^4$ theory has been studied by a variety of 
methods. Lattice models have been studied by using high-temperature (HT)
expansions and Monte Carlo (MC) simulations. Field theoretic approaches 
are the $\epsilon$-expansion and the perturbative expansion in $D=3$ fixed.
Accurate results were recently reported by using the functional renormalization
group (FRG) method. 
Recently, accurate estimates of critical exponents were obtained by using 
the conformal bootstrap method. 
We have summarized numerical results for the 
critical exponents $\nu$ and $\eta$ and the correction exponent $\omega$ 
for $N=4$, $5$ and $10$
in table \ref{Summary}. In particular in the case of field theoretic 
methods we are not able to cover the large number of works presented in 
the literature. We focus on recent results. For more extended surveys 
we refer the reader for example to refs. \cite{Kleinert,PeVi02}.

The authors of ref. \cite{Butera} computed the HT expansion coefficients
of the magnetic susceptibility and the second moment 
correlation length as rational functions of $N$
for the $O(N)$ invariant model $\phi^4$ model in the limit $\lambda=\infty$
on the simple cubic and the body centered
cubic (bcc) lattice up to the order $\beta^{21}$. They have analyzed the series
by using inhomogeneous differential approximants (DA). In ref. \cite{Butera}
they give numerical estimates for the inverse critical temperature $\beta_c$ 
and the critical exponents $\nu$ and $\gamma=\nu (2-\eta)$ for 
$N=0$, $1$, $2$, $3$, $4$, $6$, $8$, $10$, and $12$.
They give estimates based on an unbiased
analysis and an analysis that takes into account a leading correction with 
the exponent $\theta=\omega \nu$, where the values of $\theta$ are taken
from field theory. In table  \ref{Summary} we report only results obtained
for the simple cubic lattice. Those obtained for the bcc lattice are similar.
Note however that the results of the
unbiased and the biased analysis differ by more than the error bars that
are quoted. The estimates for the critical exponents essentially agree
with ours. However the error is clearly larger than ours.
The same observation holds for the estimates of $\beta_c$. It would
be an interesting exercise to perform a biased  analysis of the series
by using our values of $\beta_c$.

Let us turn to Monte Carlo simulations of lattice models.
In ref. \cite{Liu12} an  $O(N)$-symmetric loop model has been simulated
for $N=0$, $0.5$, $1$, $1.5$, $2$, $3$, $4$, $5$ and $10$.
The estimates for the critical exponents $\nu$ and $\eta$ are consistent with
but less precise than ours. In refs. \cite{Deng06,ourO5} the $\lambda=\infty$
limit of the model studied here has been simulated for $N=4$ and $5$,
respectively. The estimates for the critical exponents $\nu$ and $\eta$ are 
consistent with but less precise than ours. 
In refs. \cite{O34,O234}, similar to 
the present work, the $\phi^4$ model for $N=4$ is studied for various values of 
$\lambda$. Also here we find that the estimates for the critical exponents 
are consistent with but less precise than ours.

The authors of ref. \cite{Kos15} give rigorous error bars. Indeed our estimates
of $\nu$ and $\eta$ for $N=4$ 
are within the range allowed by the result of ref. \cite{Kos15}. The numbers taken
from table 2 of ref. \cite{CaHaSe16} have a plausible but not rigorous error bar.
In the case of the exponent $\nu$ the estimate agrees with ours within the quoted
error. In contrast, the estimate of $\omega$ is by roughly twice the error bar larger than 
ours. Note that similar observations hold when comparing the results of \cite{CaHaSe16}
for $N=2$ and $3$ with refs. \cite{myClock,myIco}.

The $\epsilon$-expansion has been extended recently to 6-loop  \cite{KoPa17} 
and to 7-loop \cite{Sch18}.  In order to get a numerical result for 
$4-\epsilon=d=3$ a resummation of the series is needed. In the literature
one can find a number of different estimates based on the 5-loop series.
In particular the estimates of the errors strongly vary. It is beyond 
our expertise to discuss the different approaches and their respective 
merits. Here we just like to remark that the estimate of $\nu$
for $N=4$ given in ref. \cite{Sha21} clearly differs from our estimate.
The same holds for the estimate of $\omega$ for $N=4$ given in ref. 
\cite{KoPa17}.

Throughout we see a good agreement with the results of ref. \cite{DePo20}.
In the case of $\eta$ our results are considerably more accurate than 
those of ref. \cite{DePo20}.

\begin{table}
\caption{\sl \label{Summary}
We summarize results for the critical exponents $\nu$ and $\eta$ and the 
exponent $\omega$ of the leading correction given in the literature.
These results were obtained from high-temperature (HT) expansions and
Monte Carlo (MC) simulations of lattice models, the conformal bootstrap
(CB) method,
the $\epsilon$-expansion,
the perturbative expansion in $d=3$ and the functional renormalization 
group.
In the case of ref. \cite{Butera} the authors give estimates of 
the  exponents $\nu$ and $\gamma$. Here we have computed $\eta=2-\gamma/\nu$.
Since it is unclear how the error propagates, we abstain from quoting one.
For a discussion see the text.
}
\begin{center}
\begin{tabular}{lllllll}
\hline
$N$ &method &year  & Ref  &    $\nu$   & $\eta$  & $\omega$  \\
\hline
4 & HT  &1997  & \cite{Butera} & 0.750(3)   & 0.0347  &    -  \\
4 & HT $\theta$-biased&1997  & \cite{Butera} & 0.759(3)  & 0.0356 &  -\\
4 & MC  &2001  &\cite{O34}    & 0.749(2)  & 0.0365(10) &     -   \\
4 & MC  &2006  &\cite{Deng06} & 0.7477(8) & 0.0360(4)  &     -  \\ 
4 & MC  &2011  & \cite{O234}   & 0.750(2)  & 0.0360(3)  &     -   \\
4 & MC  &2012  & \cite{Liu12}  & 0.7508(39) & 0.034(4)   &    -  \\
4 & MC  &2021  & this work    & 0.74817(20) &0.03624(8)  & 0.755(5)   \\
4 & CB  &2015  &  \cite{Kos15} & 0.7472(87) &  0.0378(32)  & - \\
4 & CB  &2016  &  \cite{CaHaSe16} & 0.7508(34) & -  & 0.817(30) \\
4 & $\epsilon$, 5-loop &  1998 & \cite{GuZi} & 0.737(8) & 0.036(4) & 0.795(30) \\
4 & $d=3$-exp &1998 &\cite{GuZi}  & 0.741(6) & 0.0350(45) & 0.774(20) \\
4 & $\epsilon$, 6-loop & 2017 & \cite{KoPa17} & 0.7397(35) & 0.0366(4) & 0.794(9)  \\
4 & $\epsilon$, 7-loop & 2021 & \cite{Sch18,Sha21}  & 0.74425(32)& 0.03670(38)& 0.7519(13) \\
4 & FRG &2020  &\cite{DePo20} & 0.7478(9)  & 0.0360(12) &  0.761(12) \\
\hline
5 &MC &2005 & \cite{ourO5}  & 0.779(3)  & 0.034(1)   &    -       \\
5 &MC &2012 & \cite{Liu12}  & 0.784(7)  & 0.034(6)   &    -       \\
5 & MC & 2021 & this work   &   0.7802(6) &0.03397(9) &    0.754(7)  \\
5 & FRG &2020 & \cite{DePo20} & 0.7797(9) & 0.0338(11) &  0.760(18) \\
\hline
10 & HT  &1997  & \cite{Butera} & 0.867(4)  &  0.0254 &    -  \\
10 & HT $\theta$-biased&1997  & \cite{Butera}& 0.894(4) & 0.0280 & - \\
10 & MC & 2012 & \cite{Liu12}  & 0.876(12)   &0.028(6)   &    -      \\
10 & MC & 2021 & this work     & 0.8797(9)   &0.02302(12)& 0.816(16) \\
10 & FRG &2020 & \cite{DePo20} & 0.8776(10)  &0.0231(6) &  0.807(7) \\
\hline
\end{tabular}
\end{center}
\end{table}

In table \ref{Summary} we have taken $N=4$, $5$ and $10$ as examples.
Our results for the critical exponents for $N=6$, $8$, and $12$ can be found 
in table \ref{exponents}. In the Appendix \ref{largeNinter} we interpolate
our results by using Pad\'e approximants of extended large $N$ series.

\section{Acknowledgement}
This work was supported by the Deutsche Forschungsgemeinschaft (DFG) under 
the grants HA 3150/5-1 and HA 3150/5-2.

\appendix
\section{Fits with a free parameter on the left hand side of the equation}
\label{strangefit}
We study the improved magnetic susceptibility and improved slopes, eqs.~(\ref{chiimp},\ref{Simp1}). 
These can be written as
\begin{equation}
\label{impAnsatz}
 \bar{Y}_{imp} = \bar{U}_4^x  \bar{Y},
\end{equation}
where $x$ should be tuned such that leading corrections to scaling are eliminated and
$Y$ represents either the  magnetic susceptibility or a slope. To this end we intend to perform
a fit with $x$ as free parameter.
\begin{equation}
\label{General_Ansatz}
    \bar{Y}_{imp}(x,L,\lambda) = A(L,\lambda,\{P\}) \;,
\end{equation}
where the Ansatz $A(L,\lambda,\{P\})$ is given for example by 
eqs.~(\ref{chiansatz0},\ref{chiansatz1},\ref{chiansatz2}) in the 
case of the magnetic susceptibility and $\{P\}$ is the set of free parameters
of the Ansatz. In particular, $A(L,\lambda,\{P\})$ should not 
contain terms that represent the leading correction to scaling.
We intend to perform the fit by using the function \verb+optimize.curve_fit()+
of the optimize package of python. The problem is that the parameter $x$ is 
on the left side of eq.~(\ref{General_Ansatz}).
In order to deal with this problem, we divide eq.~(\ref{General_Ansatz}) by $\bar{Y}_{imp} $ on 
both sides of the equation.  Now we treat $\bar{U}_4$ and $\bar{Y}$ along with $L$ as $X$ and 
the value of $y$ is equal to $1$.  As statistical error $\epsilon(y)$ of $y$ we assume  
$\epsilon(y)=\epsilon(\bar{Y}_{imp})/\bar{Y}_{imp}$.  In order to determine the  statistical error of 
$\bar{Y}_{imp}$, we take into account the covariance of $\bar{Y}$ and 
$\bar{U}_4$. It remains the 
problem that $x$ is not know a priori. Therefore we proceed iteratively. 
First the error is computed for an initial guess 
of $x$, then for the first result of the fit. Typically we get a stable result after
a few iterations.
Computing the statistical error of $\bar{Y}_{imp}$, 
we have neglected that $x$ has a statistical error. 
Therefore, in general, we regard the approach as an ad hoc approach.
In the analysis of our data for $N=4$ and $5$, 
we have benchmarked the results for 
critical exponents obtained by using the procedure discussed here with results obtained
from standard fits of data for $\lambda$-values close to $\lambda^*$. Therefore we 
regard the estimates of the error obtained here as reliable.

\section{Interpolation with large $N$}
\label{largeNinter}
The critical exponents $\nu$ and $\eta$ and the correction exponent $\omega$
have been computed by using the large $N$ expansion 
\cite{Vasil,Kondor,Abe,BrGrKr96}.
Here we give the series as collected in chapter 20 of 
the book \cite{Kleinert}:
\begin{eqnarray}
\eta &=& \frac{8}{3 \pi^2} \frac{1}{N} - \frac{8}{3} 
 \left(\frac{8}{3 \pi^2}\right)^2  \frac{1}{N^2}
    - \left[\frac{797}{18}- \left(27 \log(2)-\frac{61}{4} \right) \zeta(2) +
     \frac{189}{4} \zeta(3) \right] 
 \left(\frac{8}{3 \pi^2} \right)^3 \frac{1}{N^3}  + O(N^{-4}) \nonumber \\
 &=& 0.27018982305  \frac{1}{N}-0.1946734413  \frac{1}{N^2} -1.8812345072 \frac{1}{N^3} + O(N^{-4}) \;,
\end{eqnarray}
where we have evaluated the coefficients to get a better idea of their 
magnitude. The exponent of the correlation length
\begin{eqnarray}
\nu &=& 1 - 4 \frac{8}{3 \pi^2}  \frac{1}{N}   
 -  \left[\frac{9}{2} \pi^2 -\frac{56}{3} \right] \left(\frac{8}{3 \pi^2}\right)^2  \frac{1}{N^2}
+ O(N^{-3})  \nonumber \\
&=& 1 - 1.0807592922  \frac{1}{N} -  1.8795637876 \frac{1}{N^2}  + O(N^{-3})
\end{eqnarray}
and correction exponent
\begin{eqnarray}
\omega &=& 1 - 8 \frac{8}{3 \pi^2}  \frac{1}{N} -
2 \left[\frac{9}{2} \pi^2 - \frac{104}{3}\right] 
 \left(\frac{8}{3 \pi^2}\right)^2   \frac{1}{N^2}
+ O(N^{-3}) \nonumber \\
&=& 1  - 2.1615185844 \frac{1}{N} - 1.4230462800 \frac{1}{N^2} + O(N^{-3})
\;.
\end{eqnarray}
As stated in the literature, evaluating the series naively or 
by using Pad\'e approximants, numerically useful results can be expected
at best for $N \gtrapprox 10$.  

Here we like to extend the series by one or two orders, where the additional
coefficients are determined by fitting the numerical results for 
$N=4$, $5$, $6$, $8$, $10$, and $12$ obtained here with  Pad\'e approximants 
of the extended series. We are aiming at a reasonable 
interpolation for $N=7$, $9$, $11$, and for $N$ somewhat larger than $12$. 
We used a standard $\chi^2$ minimization. Interpreting the result, one has
to keep in mind that the error 
that we quote for the exponents is partially of systematic nature.

In the case of $\eta$ we get an acceptable fit down to $N=8$ for adding a
$c_4 N^{-4}$ term and a $[1,3]$ Pad\'e approximant. As result for the 
coefficient we get $c_4=-4.78(84)$. Adding a $c_4 N^{-4}$  and a $c_5 N^{-5}$
term, we get acceptable
fits down to $N=5$ by using $[0,5]$ or $[3,2]$ approximants.
For the $[1,4]$ Pad\'e approximant we get an acceptable fit even down to $N=4$.
The estimates of $c_4$ and in particular of $c_5$ differ considerably between
the different Pad\'e approximants that we used.
The estimates of $c_4$ and $c_5$ and the associated covariance matrices are
contained in a Python3 script 
that we provide as supplemental material.
This Python script computes $\eta$ for $N \ge 5$ based on the Pad\'e 
approximants discussed here.

In the case of $\nu$ and $\omega$ we performed similar fits. Also here, 
the results are given in Python3 scripts 
that produce estimates of 
$\nu$ and $\omega$ for $N \ge 5$.

\end{document}